\documentclass{aa}
\usepackage{txfonts}
\usepackage{graphicx}
\usepackage{natbib}

\bibpunct{(}{)}{;}{a}{}{,}

\usepackage{color}

\newcommand{\kms}{\,{\rm km\,s}^{-1}}

\newcommand{\Lsun}{L_\odot}
\newcommand{\mdot}{\dot{M}}
\newcommand{\Msun}{\,\mathrm{M}_\odot}

\newcommand{\as}{\ifmmode {^{\scriptscriptstyle\prime\prime}}
        \else $^{\scriptscriptstyle\prime\prime}$\fi}

\newcommand{\simless}{\mathbin{\lower 3pt\hbox
      {$\rlap{\raise 5pt\hbox{$\char'074$}}\mathchar"7218$}}} 
\newcommand{\simgreat}{\mathbin{\lower 3pt\hbox
     {$\rlap{\raise 5pt\hbox{$\char'076$}}\mathchar"7218$}}} 

\begin{document}
\title{Faint disks around classical T Tauri stars:\\ small but dense enough to form planets
\thanks{Based on observations carried out with the IRAM Plateau de Bure interferometer.
 IRAM is supported by INSU/CNRS (France), MPG (Germany) and IGN (Spain).}
}
%

%
\author{V. Pi\'etu \inst{1}, S. Guilloteau \inst{2,3}, E. Di Folco \inst{2,3}, A. Dutrey\inst{2,3}
and Y. Boehler\inst{4}}
\institute{
IRAM, 300 rue de la piscine, F-38406
Saint Martin d'H\`eres, France
\and
Univ. Bordeaux, LAB, UMR 5804, F-33270, Floirac, France
\and
CNRS, LAB, UMR 5804, F-33270 Floirac, France\\
  \email{[guilloteau,dutrey]@obs.u-bordeaux1.fr}
\and Centro de Radioastronom\`ia y Astrof\`isica, UNAM, Apartado Postal 3-72, 58089 Morelia, Michoac\`an, Mexico
}

\offprints{S.Guilloteau, \email{Stephane.Guilloteau@obs.u-bordeaux1.fr}}

\date{Received / Accepted } %
\authorrunning{Pietu et al.} %
\titlerunning{Faint disks around classical T Tauri stars} 

\abstract
{
Most Class~II sources (of nearby star forming regions) are surrounded by disks with weak 
millimeter continuum emission. These ``faint'' disks may hold clues to the disk dissipation
 mechanism. However, the physical properties of proto-planetary disks have been directly 
 constrained by imaging only the brightest sources.
}
{
We attempt to determine the characteristics of such faint disks around classical T Tauri stars, and
to explore the link between disk faintness and the proposed disk dispersal mechanisms
(accretion, viscous spreading, photo-evaporation, planetary system  formation).
}
{
We performed high-angular resolution ($0.3''$) imaging of a small sample of disks (9 sources) 
with low 1.3\, mm continuum flux (mostly $<30$\,mJy) with the IRAM Plateau de Bure interferometer 
and simultaneously searched for $^{13}$CO (or CO) J=2-1 line emission.
Using a simple parametric disk model, we determine characteristic sizes of the disks, in dust and gas,
and we constrain surface densities in the central 50\,AU.
}
{
All disks are much smaller than the bright disks imaged so far, both in continuum and $^{13}$CO lines (5 detections).
In continuum, half of the disks are very small, with characteristic radii less than 10\,AU, but still 
have high surface density values.
Small sizes appear to be the main cause for the low disk luminosity. Direct evidence for grain growth
is found for the three disks 
that are sufficiently resolved. Low continuum opacity is attested in two systems only, but we cannot
firmly distinguish between a low gas surface density and a lower dust emissivity resulting from grain growth.
Finally, we report a tentative discovery of a $\sim 20$\,AU radius cavity in DS Tau, which with the
(unresolved) ``transition'' disk of CX~Tau, brings the proportion of ``transitional'' disks to a similar
value to that of brighter sources. The existence of cavities cannot explain by itself their observed low mm flux.
}
{
This study highlights a category of very compact dust disks, still exhibiting high surface densities, 
which may represent up to 25\% of the whole disk population. While its origin is unclear with the
current data alone, it may be related to the compact planetary systems found by the Kepler
mission.}


\maketitle{}

\section{Introduction}

The many known exoplanet systems display tremendous variety in the masses of their
stars, the masses of planets they host, and their architectures.  Study of the early evolution of
protostellar disks will contribute to our understanding of the origins of this diversity. "Faint"
disks (which we defined as having 1.3 mm continuum flux $\simless 60$ mJy at the Taurus distance, 140 pc)
play a key role in this respect.
First of all, most disks around Class II sources are ``faint''. \citet{Andre+Montmerle_1994}
find that 50 \% of the Class II sources in $\rho$ Oph (at 120 pc) have 1.3\,mm flux $< 40$ mJy, while
only 25 \% have flux $> 125$~mJy. Similar statistics were obtained in the Taurus region by \citet{Beckwith+etal_1990}.
Second, young stars of mass  $< 0.3$\,M$_\odot$ (the most numerous in our Galaxy) are expected to be surrounded by 
low-mass, intrinsically faint disks from theoretical considerations. Until the work of \citet{Andrews+etal_2013},
this trend was not clearly observed because of incompleteness of the samples towards low masses (sensitivity). The
large dispersion of disk continuum brightness as a function of stellar mass also loosen the expected correlation.
Finally, low-mass stars with faint disks play a critical role in validating stellar evolution models. Providing 
accurate stellar masses in the 0.2-0.4\,$\Msun$ range would allow critical tests of early stellar evolution models, 
hopefully offering better age estimates for all young stars to ultimately establish a more reliable clock for 
planetary system formation.

Disks can be ``faint'' for many different reasons. \textit{Low disk surface density} is the simplest reason. 
Temperature also affects continuum flux, but varies over a much more limited range. \textit{Dust settling} 
when seen at high inclination  \citep[$> 80^\circ$, e.g. the edge-on objects HK Tau B, HV Tau C in][]{Guilloteau+etal_2011} 
leads to smaller flux \citep[see the study of][]{Boehler+etal_2013}, as the warmer parts are hidden by opacity.  
Disks may also \textit{just be small}, as for example \object{BP Tau} \citep{Dutrey+etal_2003}, sometimes as 
a result of \textit{outer disk truncation}, as happens in binaries \citep[e.g. Haro 6-10, UY Aur,][]{Guilloteau+etal_2011,Duvert+etal_2000}.  On another hand, \textit{inner disk clearing}, which suppresses the densest parts of the disk, is also observed in many objects \citep[e.g. AB Aur, LkCa 15, MWC 758,][]{Pietu+etal_2005,Pietu+etal_2006,Isella+etal_2010a}. Finally, \textit{grain growth}, which reduces the opacity per unit mass, can lead to small (sub)mm continuum flux densities.

These causes are not mutually exclusive: \object{HH 30} \citep{Guilloteau+etal_2008} combines small size, inner 
hole and grain growth, as well as a somewhat lower than average surface density. The various possible causes 
have different observational consequences.  \citet{Ricci+etal_2010a} found
a correlation between the spectral index $\alpha$ ($S(\nu) \propto \nu^\alpha)$ with S(1.3\,mm), with 
$\alpha \simeq 2$ for low flux densities. $\alpha = 2$ indicates either optically thick emission, or 
large grains (``pebbles''), as the dust emissivity index $\beta = \alpha-2$ in the optically thin case. 
However, the observed correlation fails to be reproduced by current models of disk evolution including 
grain growth and viscous evolution \citep[][their Fig.2]{Birnstiel+etal_2010}. A trend for smaller sizes 
at low flux \citep[][their Fig.7]{Andrews+etal_2010} supports the compact disk interpretation.  
Furthermore, since \citet{Guilloteau+etal_2011} showed that larger grains are preferentially found
in the inner 60 - 100 AU of disks, small disks are also expected to have lower $\alpha$, producing 
the observed correlation between flux and spectral index.

Whatever the cause of the faintness, ``faint'' disks are essential objects for understanding
the planetary system formation process. Grain growth and dust settling are major steps in this respect.
Inner cavities can be signposts of planetary system formation, while disks with low surface density may be in the dust
(and gas) dissipation stage. To separate between the possible causes, resolved images are essential.
Because of the resolution and sensitivity limitations of the existing mm arrays, most studies have focussed
on the brighter objects \citep[e.g.][]{Andrews+Williams_2007,Isella+etal_2009,Guilloteau+etal_2011}, so that
resolved images of faint disks are rare. A first attempt was performed by \citet{Andrews+etal_2010}, who
studied the continuum emission from some disks in the $\rho$ Oph regions,  6 of which would qualify as 
``faint'', showing that on average these disks appeared indeed smaller than brighter ones. However, 
the sensitivity was insufficient to study the
CO line emission.

We report here on a high angular resolution (0.4-0.6$''$) study in line and continuum emission of 9 
disks in the Taurus complex, 8 of them having $S_\nu < 30$ mJy at 1.3\,mm, and the last one, CW Tau, 
being at the boundary between ``faint'' and ``bright'' disks with $S_\nu \approx 60$ mJy.

\section{Observations and Data Analysis}
\label{sec:obs}

\subsection{Source Sample}
\label{sec:sub:sample}

Our sample consists in the eight disks from the \citet{Ricci+etal_2010a} study
which had no previous high angular resolution data: \object{CW Tau}, \object{CX Tau}
\object{DE Tau}, \object{DS Tau}, \object{FM Tau}, \object{FZ Tau}, \object{HO Tau} and \object{SU Aur}.
All of them have low 1.0\,mm fluxes, ranging
from 19 mJy to 70\,mJy, with the exception of CW\,Tau which is substantially brighter (129\,mJy).
We also observed \object{V836 Tau}, one of the rare stars in an intermediate stage between
cTTS and wTTS \citep{Duvert+etal_2000,Najita+etal_2008}.
Table \ref{tab:prop} summarizes the properties of the selected stars.
Note that there exist substantial disagreement between the stellar properties
reported by several authors, in particular for CW Tau and DS Tau, for which \citet{Ricci+etal_2010a}
indicate an unusually large age of 17 Myr. Table \ref{tab:prop} gives the stellar properties derived 
by \citet{Bertout+etal_2007} and \citet{Andrews+etal_2013} using the \citet{Siess+etal_2000}
evolutionary tracks.

\begin{table*}
\caption{Stellar properties}
\label{tab:prop}
\begin{tabular}{c ll | c cccccc cc c}
 Source & RA & Dec. & Spectral  &  \multicolumn{2}{c}{$L_*$} & \multicolumn{2}{c}{$M_*$} &
 \multicolumn{2}{c}{Age} & $\dot{M}$  & Ref. \\
 & \multicolumn{2}{c|}{J2000} & Type & \multicolumn{2}{c}{($\Lsun$)} &  \multicolumn{2}{c}{($\Msun$)} & \multicolumn{2}{c}{Myr} & ($\Msun$/yr) \\
 & & &  & (A) & (B) &  (A) & (B)  & (A) & (B)  &   &  \\
\hline
FM Tau &  04:14:13.591 &  28:12:48.90 & M0 & \multicolumn{2}{c}{0.25 -- 0.40} & \multicolumn{2}{c}{0.64 -- 0.60} &
\multicolumn{2}{c}{12 -- 3.4} & -8.87 & b,c\\
CW Tau &  04:14:17.011 &  28:10:57.51 & K3 & \multicolumn{2}{c}{0.68 -- 2.4\phantom{0}} & \multicolumn{2}{c}{1.11 -- 1.6\phantom{0}}  & \multicolumn{2}{c}{16 -- 2.2} & -7.99  & a \\
CX Tau &  04:14:47.869 &  26:48:10.73 & M2.5 & \multicolumn{2}{c}{0.56 -- 0.37} & \multicolumn{2}{c}{0.40 -- 0.36}  & \multicolumn{2}{c}{1.3 -- 1.7} & -9.43  & a \\
DE Tau &  04:21:55.644 &  27:55:05.91 & M1-M2 & \multicolumn{2}{c}{1.14 -- 1.03} & \multicolumn{2}{c}{0.66 -- 0.40} &  \multicolumn{2}{c}{$<$1 -- 0.9} & -8.18 & b,c \\
FZ Tau &  04:32:31.768 &  24:20:02.80 & M0 & \multicolumn{2}{c}{0.51 -- 1.25} & \multicolumn{2}{c}{0.70 -- 0.25} & \multicolumn{2}{c}{2.5 -- 0.8} & -7.7 & c \\
HO Tau &  04:35:20.218 &  22:32:14.34 & M0.5 & \multicolumn{2}{c}{0.17 -- 0.13} & \multicolumn{2}{c}{0.67 -- 0.56} & \multicolumn{2}{c}{17 -- 13} & -8.87 & b \\
DS Tau &  04:47:48.598 &  29:25:10.92 & K5 & \multicolumn{2}{c}{0.66 -- 0.76} & \multicolumn{2}{c}{1.04 -- 1.04} & \multicolumn{2}{c}{7 -- 4} & -7.39 & b \\
SU Aur &  04:55:59.392 &  30:34:01.23 & G2 & \multicolumn{2}{c}{\phantom{0}9.3 -- 10.7} & \multicolumn{2}{c}{1.9 -- 2.5} & \multicolumn{2}{c}{6.3 -- 2.9} & - & a \\
V836 Tau &  05:03:06.596 &  25:23:19.59 & K7 & \multicolumn{2}{c}{0.58 -- 0.58} & \multicolumn{2}{c}{0.7\phantom{0} -- 0.83} &  \multicolumn{2}{c}{\phantom{0}1 -- 3.7} & -9.0 & d\\
\hline
\end{tabular}
\tablefoot{Spectral types, and for (A) also luminosity and ages are from
(a) \citet{Bertout+etal_2007}, (b) \citet{Kenyon+Hartmann_1995},
(c) \citet{White+Ghez_2001}, (d) \citet{White+Hillenbrand_2004};
(B) luminosity and ages from \citet{Andrews+etal_2013} using the
\citet{Siess+etal_2000} tracks.
Accretion rates are from  \citet{White+Ghez_2001},
except for V836 Tau \citep{Najita+etal_2008}. }
\end{table*}

\subsection{Observations}
\label{sec:sub:obs}
Observations of the main sample were carried out with the IRAM Plateau de Bure interferometer, using
the ABC configuration. The dual-polarization receivers were tuned near
220 GHz to cover the J=2-1 transitions of $^{13}$CO and C$^{18}$O.
The equivalent wavelength for the continuum is 1.36\,mm.

The C configuration was observed on Dec 14, 2010, the B
configuration on Feb 13 and Mar 5, 2011 and the A configuration on Feb 20, 2013. Typical system temperatures
were 90 -- 150 K (0.5 mm of precipitable water vapor only) , 150 -- 180 K,
and 200 -- 350 K resp\., with scattered clouds on Mar 5, and 90 -- 150 K for the A configuration.
The rms phase noise was of order $30^\circ$ for the C configuration, $20 - 50^\circ$ for the A
configuration, but rather
high ($50 - 70^\circ$) for the B configuration, resulting in a limited seeing.
All 8 sources were observed in snapshot mode, providing limited but well distributed
UV coverage. This mode provides a uniform calibration for all sources. We obtain
about 2 hrs on source integration time for each source. The typical angular
resolution is $0.6 \times 0.4''$ at PA$\sim 30^\circ$ with either robust or natural
weighting.

The narrow band correlator was set to cover the CO isotopologues with 0.05 $\kms$
channel separation, later smoothed to $0.21 \kms$ spectral resolution.
The typical point source sensitivity is 20 mJy per channel, corresponding to a brightness
of 1.6 K.

The wideband correlator covered 4 GHz in each polarization,
leading to a typical sensitivity of 0.13 mJy/beam.
Phase noise however limits the dynamic range, especially on the brighter sources.
These sources are compact and strong enough: we used one iteration of
phase self-calibration, using an integration time of 120 sec,
bringing the effective image plane noise close to the theoretical value
for the BC configuration. The self-calibration initial model retained
only the 10 first Clean components of the source: this is essentially equivalent
to assuming the source is centro-symmetric and solve for the antenna phases under
this assumption.
For the fainter sources, the typical rms noise per antenna is 1.4 to 3.5 mJy (depending
on the day) for a 120 sec integration timescale, which in principle yields an antenna
based rms of 9 to 25 degrees in the self calibration process.  Phase fluctuations
at timescales between 45 sec (the dump time used in these observations) and the 120
sec integration time of the self calibration process are not corrected.  We checked that
using a shorter integration time for the self calibration process
only brings limited improvements for the brighter
sources, and in particular does not significantly affect the derived disk parameters.

The IRAM Plateau de Bure interferometer is equipped with 22 GHz radiometers that
correct for phase errors on timescales from 1 sec to 45 sec, the current integration
time per dump, leaving essentially Gaussian phase errors on these timescales with
a mean rms around 30$^\circ$. This first phase correction results in an essentially
constant decorrelation factor, which is compensated by the amplitude calibration on the
nearby quasars. The flux calibration was performed in comparison with MWC\,349, which
has a flux density of 1.9 Jy at 220 GHz. As some sources are bright enough, the
consistency of the amplitude calibration between the different days could be checked,
and was found to be on the order of 5 \%.

The phase corrections derived by self-calibration of the continuum data were
transferred to the $^{13}$CO data. Positions of the detected sources
are given in Table \ref{tab:prop}, with an astrometric accuracy of order 0.05$''$.
No proper motion corrections were applied, but the use of self-calibration
implicitly correct for any source motion over the 2 years observing period.

The V836 Tau data were obtained separately (Project N035). Long baseline (B configuration)
data was obtained on 27 Feb and 3 Mar 2004, and the C configuration was observed on 24 Nov 2004.
The single polarization, dual frequency receivers were tuned to the CO J=2-1 and J=1-0 frequencies.
These data were merged with the compact configuration data from \citet{Duvert+etal_2000} obtained
in 1997 Sep 28 and Oct 29. The effective rms noise for the continuum is 0.5 mJy/beam.
Since these data only have 500 MHz of effective bandwidth (instead of 8 GHz for the other sources),
we  used a longer timescale of 480 s for the self calibration.

\begin{figure*}
   \includegraphics[width=15.0cm]{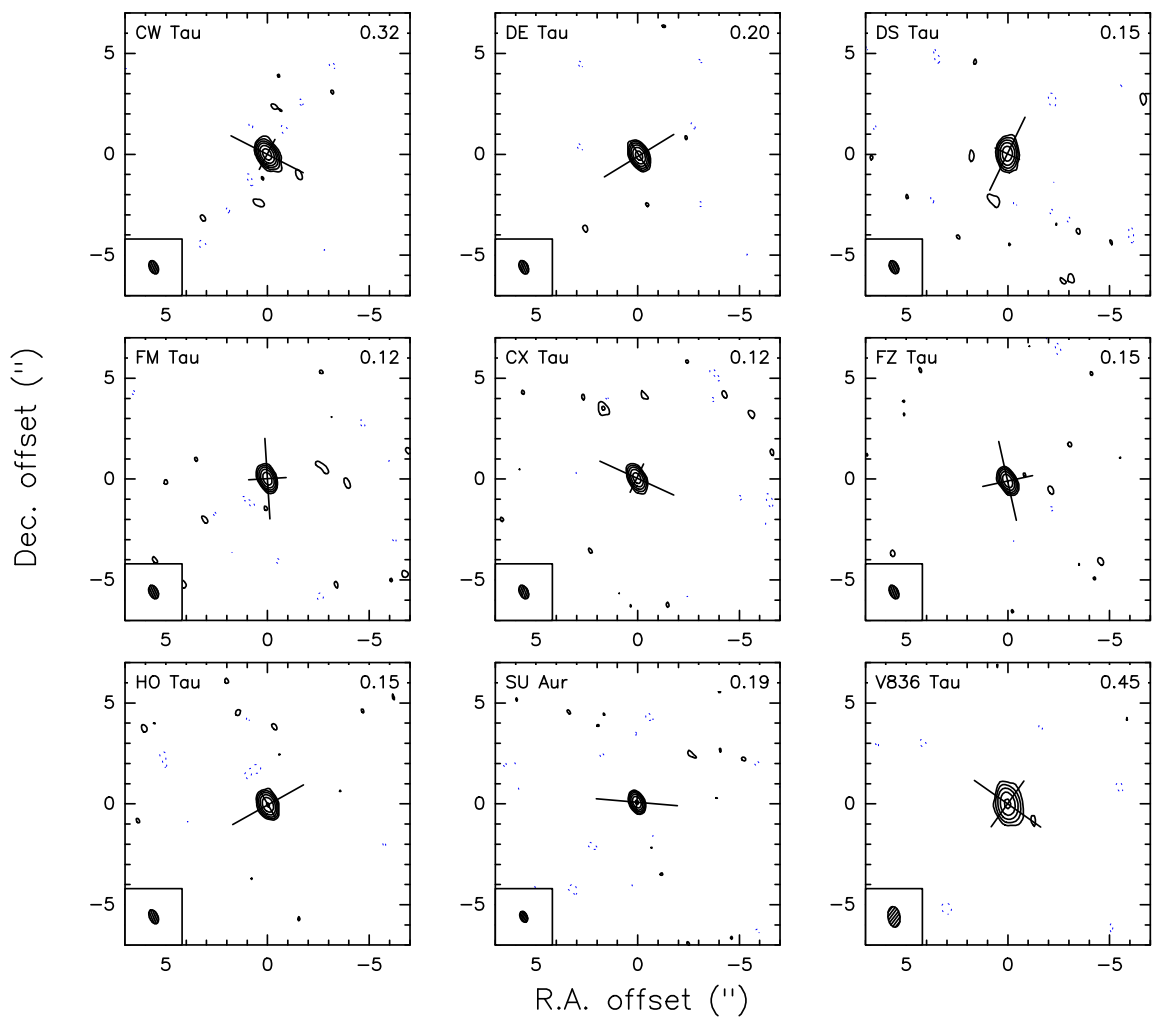}
 \caption{Naturally weighted continuum images for all sources. Contour levels are
 -3, 3, 6 12, 24, 48, and 96 times the noise level. Noise level (in mJy/beam)
  is indicated in the upper right corner of each panel.
Only CX Tau has a potential secondary source: a 6 $\sigma$ signal (0.7 mJy) is located
approximately 3.8$''$ NE of CX Tau. Positions are relative to the
coordinates in Table \ref{tab:prop}.
  }
  \label{fig:large}
\end{figure*}

\begin{figure*}
   \includegraphics[width=15.0cm]{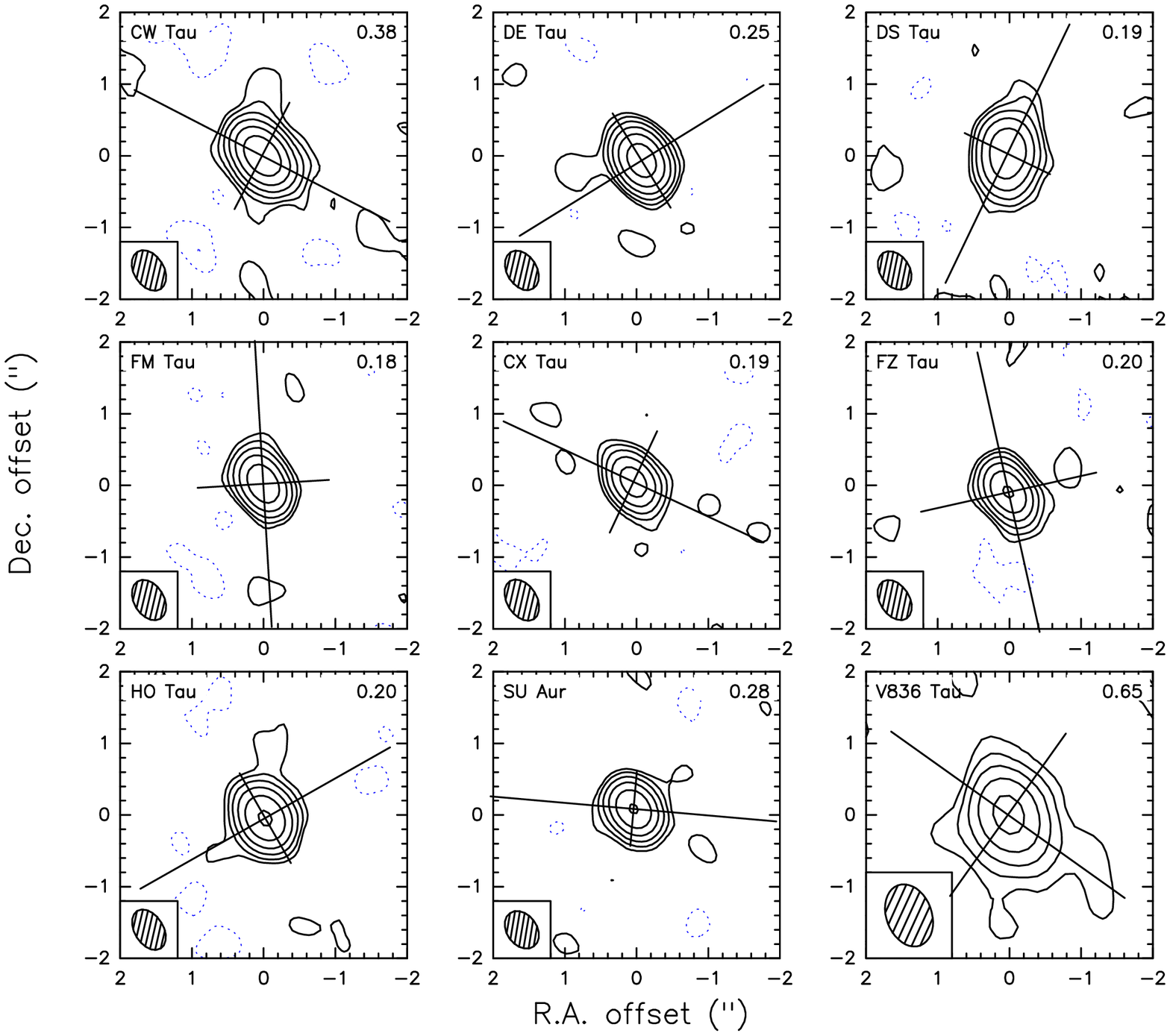}
 \caption{Continuum images obtained with robust weighting. Contour levels are
 -4, -2, 2, 4, 8, 16, 32, 64, and 128 times the noise level. Noise level (in mJy/beam)
  is indicated in the upper right corner of each panel.  Crosses indicate the
  orientation and aspect ratio of the disks derived from model fitting.
  Positions are relative to the coordinates in Table \ref{tab:prop}.}
  \label{fig:small}
\end{figure*}

Figure \ref{fig:large} shows the images obtained with natural weighting after self-calibration.
All disks are well detected, and at this scale, appear very compact with essentially no significant
residual emission in the field of view, except for two faint (0.7 mJy) point sources, one 4$''$ North of CX Tau
and another 7$''$ NW of DS Tau. An enlarged view, using images generated with
robust weighting to enhance the angular resolution, is shown in Fig.\ref{fig:small}. Even
at this scale, the sources appear compact, and furthermore, convolution by the elongated
beam masks the intrinsic elongated aspect expected from inclined disks.

\subsection{Analysis Method}

We use the DiskFit tool \citep{Pietu+etal_2007} to fit a parametric
model of flared disks to the calibrated visibilities, in the same way as \citet{Guilloteau+etal_2011}.
For the continuum, since we only have images at one frequency, we used a number of
simplifying assumptions:
\begin{itemize}
\item a power law for the temperature
\begin{equation}\label{eq:temp}
    T(r) = T_{0} \left(\frac{r}{R_0}\right)^{-q} ,
\end{equation}
with $T_0 = 15$ K at $R_0 = 100$ AU and an exponent $q = 0.4$
\item uniform dust properties, with a dust absorption coefficient
    $\kappa_\nu$(220 GHz) = 2 cm$^{2}$g$^{-1}$,
     and an opacity scaling as $\nu^\beta$ with frequency. We
     also assume a gas to dust ratio of 100.
\item two different  surface density distributions, a simple truncated power law (Model 1)
\begin{equation}\label{eq:power}
    \Sigma_g(r) = \Sigma_{0} \left(\frac{r}{R_0}\right)^{-p} ,
\end{equation}
for $R_\mathrm{int} < r < R_\mathrm{out}$,
and a self-similar viscous law (Model 2)
\begin{equation}\label{eq:edge}
\Sigma_g(r) = \Sigma_0 \left(\frac{r}{R_0}\right)^{-\gamma}  \exp\left(-(r/R_c)^{2-\gamma}\right) .
\end{equation}
for $R_\mathrm{int} < r$.
\item a flared geometry with a scale height
\begin{equation}\label{eq:height}
    h(r) = H_{0} \left(\frac{r}{R_0}\right)^{h} ,
\end{equation}
with $H_0 = 16$ AU, and $h=1.25$.
\end{itemize}
The disks are then further characterized by the system geometry: position ($x_0,y_0$), inclination $i$ and
orientation $PA$. We follow the convention of \citet{Pietu+etal_2007} and indicate the PA of the
disk rotation axis.
The limited compromise in sensitivity and angular resolution precludes obtaining good independent
constraints on the coupled parameters $(p,R_\mathrm{out})$ in Model 1, or ($\gamma,R_c$) in
Model 2. We thus assumed $p = 1$ for all sources in Model 1, and $\gamma = 0.5$ in Model 2, so that
the outer radius  $R_\mathrm{out}$ or the critical radius $R_c$ become the sole indicator of the disk
compactness.
The validity of these assumptions was checked on the brightest source, CW Tau, where we
found $p \approx 1.2 \pm 0.4$. The high resolution images were complemented by flux measurements at 0.85 mm
\citep[from][]{Andrews+Williams_2005} and near 3\,mm \citep[from][]{Ricci+etal_2010a} to allow
a derivation of the dust emissivity index. The arbitrary choice of $p$ or $\gamma$ has
limited impact on the derived sizes.
In Model 1, $R_\mathrm{out}$ would typically be $\sim 15$ \% smaller for $p=0$, and $\sim 5 \%$ larger for
$p=2$. In Model 2, using $\gamma = 0$ would increase $R_c$ by 20 \%, while using $\gamma \geq 1$ would
result in much smaller values, although the fit quality significantly degrades for the larger sources.

\begin{figure*}
   \includegraphics[width=15.0cm]{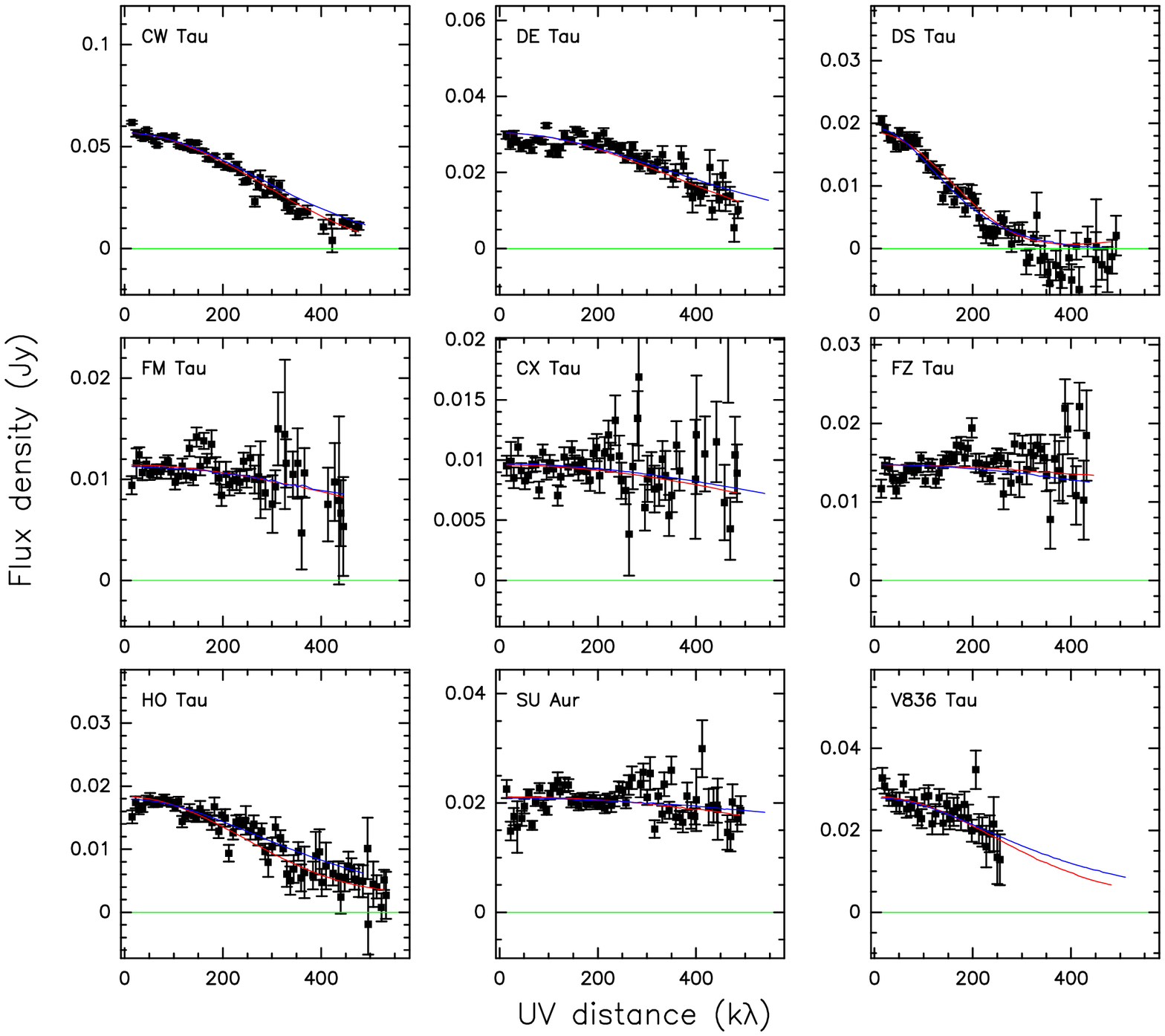}
 \caption{Deprojected visibility profiles for the sources on a common relative scale, with best-fit
 profiles superimposed. Red is for the power-law fit, blue for the
 viscous profile.}
  \label{fig:self}
\end{figure*}

Given the sparse $uv$ coverage, the analysis is done in the $uv$ plane by fitting the observed visibilities
to avoid non-linear effects due to deconvolution. The flux measurements at other wavelengths are considered
as zero-spacing visibilities: they only constrain $\beta$ (and $T_0$ for SU Aur).
Results are given in Table \ref{tab:self}. Figure \ref{fig:self} shows the visibilities and the model fit
for all sources, after deprojection from inclination effects and circular averaging. Most sources being
quite small, self-calibration is essential to measure the source size. Without self-calibration,
the phase noise on the longest baselines results in a substantial seeing limitation, making the apparent characteristic size of the sources  $R_c > 20$ AU.
More details are given in Appendix \ref{app:self}.

The errorbars in Table \ref{tab:self} were in general computed from the covariance
matrix. However, this is not appropriate for several parameters. For highly inclined objects,
the errorbar on the inclination is highly asymmetric. Similarly, for very compact sources, the errorbar
on the radii (either $R_c$ or $R_\mathrm{out}$) must be asymmetric.

For these parameters,
as well as for the position angle, we explored the $\chi^2$ curve as a function
of parameter value, fitting all other parameters, to
derive the $1 \sigma$ errorbars or $3 \sigma$ upper limits. Finally, although inclinations
are sometimes highly uncertain, this has little impact on the derived characteristic radii:
$R_\mathrm{out}$ is well constrained in all cases. These effects are described in
more details in Appendix \ref{app:errors}.

For the CO isotopologue results, we also assumed the CO surface density to be a power law.
As for continuum, the signal to noise is insufficient to constrain the exponent of
this power law: we assumed it to be $p = 1.5$, because measurements of spectral lines
in general give steeper radial dependencies for molecules than for dust \citep[e.g.][]{Pietu+etal_2007}.
The continuum data was subtracted from the line data before performing the fit, as discussed
in \citet{Pietu+etal_2007}.
The velocity field was assumed to be Keplerian:
\begin{equation}\label{eq:kepler}
    V(r) = V_{100} (r/100 \mathrm{AU})^{-0.5}
\end{equation}
and the local linewidth was set to 0.3 $\kms$. The systemic velocity was derived from the
fitting process. Orientation and inclinations were derived independently
from the continuum. Note that while these two parameters are affected by the amplitude calibration only
for the continuum, they are essentially affected by the phase bandpass calibration for the line data.
A good match between these independent derivations indicates that the calibration accuracy is
not the limiting factor. The results of the CO isotopologue model fitting are given in Table \ref{tab:coline}.
Errorbars were computed from the covariance matrix only. The $^{13}$CO disk models are well constrained for
CW Tau, and also DS Tau, but should only be interpreted as a possible solution for the weaker sources
CX Tau, DE Tau and SU Aur.

The main uncertainty in our disk model is the temperature at 100 AU, $T_0$. The $^{13}$CO results are relatively insensitive to the assumed temperature. However, concerning dust, temperature can affect the disk mass, which will scale as $1/T$ for mostly optically thin emission, or the disk size for mostly optically thick emission in unresolved sources. We expect the effect to be small: our adopted value
is reasonable given the typical disk size in our sample (see Sec.\ref{sec:tdust} for further discussion),
and we already accounted for a larger value for SU Aur.

The next critical parameter is our choice of dust absorption coefficient at 220 GHz.
Our adopted value, 2 cm$^2$g$^{-1}$, is typical for the dust grains in disks.
 \citet{Guilloteau+etal_2011} used
2 cm$^2$g$^{-1}$ (but at 230 GHz), while the dust model used
by \citet{Andrews+Williams_2007} gives 2.1 cm$^2$g$^{-1}$ at 220 GHz. However,
if grains are substantially larger than 1\,mm, the dust emissivity can be significantly reduced,
see for example Fig.5 of \citet{Isella+etal_2009} or \citet{Draine_2006}.
Thus, our reported disk masses are likely to be (rather conservative) lower limits (within the assumption of a standard gas to dust ratio).

\section{Results}
\label{sec:results}

In all cases, there is good agreement between the geometry determined from
the line data and the continuum results, which indicates that systematic effects
are negligible as the continuum is affected by amplitude calibration fluctuations, while the
spectral line data is mostly affected by bandpass calibration errors.


\begin{table*}
\caption{Continuum Results}
\label{tab:self}
\begin{tabular}{c | r rrlr cc rc}
 Source & \multicolumn{1}{c}{Flux}  &  \multicolumn{1}{c}{Inclination} & \multicolumn{1}{c}{PA} &  \multicolumn{1}{c}{$R_\mathrm{out}$}
    &  \multicolumn{1}{c}{$R_c$} & \multicolumn{1}{c}{$\Sigma_{10}$} & \multicolumn{1}{c}{Mass} & \multicolumn{1}{c}{$\beta$} & $\alpha$\\
 & \multicolumn{1}{c}{(mJy)} &  \multicolumn{1}{c}{($^\circ$)} & \multicolumn{1}{c}{($^\circ$)} & \multicolumn{1}{c}{(AU)} &  \multicolumn{1}{c}{(AU)} & \multicolumn{1}{c}{(g.cm$^{-2}$)} & \multicolumn{1}{c}{Log$_{10}$(M/M$_\odot$)} & \\
\hline
FM Tau &   11.3 $\pm$    0.2
 &  \textit{$\sim$ 70}
 &   83 $\pm$  13
 &   15 $\pm$  4 ($<28$)
 &    3 $\pm$  4
 & $60_{-50}^{+\infty}~(> 7)$ 
 & $> -3.1$ 
 &  -0.52 $\pm$   2.26
 & 2.2 $\pm$ 0.7
 \\
CW Tau &   58.7 $\pm$    0.4
 &   65 $\pm$   2
 &  332 $\pm$   3
 &   39 $\pm$  2
 &   18 $\pm$  2
 & 135
 &      -1.65 $\pm$       0.55
 &   1.29 $\pm$   0.24
 &   2.25 $\pm$   0.18
 \\
CX Tau &    9.6 $\pm$    0.2
 &   \textit{62 $\pm$  53}
 &   -5 $\pm$  23
 &   12 $\pm$  2 ($< 22$)
 &    8 $\pm$  2
 & $12_{-3}^{+8}~(> 7)$ 
 & $> -3.1$ 
 &   0.53 $\pm$   0.53
 &   2.2 $\pm$   0.6
 \\
DE Tau &   29.5 $\pm$    0.2
 &   66 $\pm$   7
 &  210 $\pm$   8
 &   19 $\pm$  3
 &   7 $\pm$  3
 & $160_{-50}^{+\infty}~(> 20)$ 
 & $> -2.5$ 
 &   1.31 $\pm$   4.41
 &   2.47 $\pm$ 0.16
 \\
FZ Tau &   14.7 $\pm$    0.1
 &   \textit{$<$ 70} 
 &  114 $\pm$   3
 &   \phantom{0}9 $\pm$  1 ($< 13$)
 &    2 $\pm$  4
 & $200_{-50}^{+\infty}~(> 20)$ 
 & $> -2.8$ 
 &  -5.15 $\pm$   5.96
 &  1.6 $\pm$ 0.5
 \\
HO Tau &   17.7 $\pm$    0.2
 &   29 $\pm$   8
 &  270 $\pm$  17
 &   48 $\pm$  3
 &   22 $\pm$  1
 & 13
 &      -2.56 $\pm$       0.16
 &   0.29 $\pm$   0.28
 &  2.1 $\pm$ 0.3
 \\
DS Tau &   20.1 $\pm$    0.3  
 &   70 $\pm$   3
 &   63 $\pm$  11
 &   70 $\pm$  4
 &   41 $\pm$  3
 & 13
 &      -2.47 $\pm$       0.01
 &   0.01 $\pm$   0.17
 &   1.61 $\pm$   0.25
 \\
SU Aur &   22.5 $\pm$    0.2   
 &   \textit{44 $\pm$  24}
 &  125 $\pm$  24
 &   13 $\pm$  2 ($< 17$)
 &    4 $\pm$  2
 & $10_{-1}^{+5}$ 
 &  $> -3.1$ 
 &   0.67 $\pm$   0.24
 &  2.62 $\pm$ 0.16
 \\
V836 Tau &   28.0 $\pm$    0.7
 &   \textit{48 $\pm$  18}
 & -122 $\pm$  28
 &   48 $\pm$  9
 &   27 $\pm$  7
 & 15
 &      -2.39 $\pm$       0.08
 &   0.39 $\pm$   0.19
 &   2.2 $\pm$ 0.3
 \\
\hline
\end{tabular}
\tablefoot{Errorbars are $1 \sigma$, and upper or lower limits
$3 \sigma$. Inclinations given in italics are rather uncertain: in this 
case, the error on position angle is only valid for sufficiently high 
inclinations. $\alpha$ is the flux spectral index, $\beta = \alpha-2$
in the optically thin and Rayleigh-Jeans limit.}
\end{table*}

\begin{table*}
\caption{Results from CO isotopologues}
\label{tab:coline}
\begin{tabular}{lc|c|c|c|c|c}
\hline
Source &                    & CW Tau       & CX Tau ($\dagger$)   & DE Tau       & DS Tau  & SU Aur(*) \\
\hline
PA (mm) & $(^\circ)$        & 332  $\pm$ 3 & -5 $\pm$ 23 & 210 $\pm$  8 &  63 $\pm$ 11  & 125 $\pm$ 24 \\
PA (CO) & $(^\circ)$        & 331 $\pm$  2 & 46 $\pm$ 10 & 255 $\pm$ 20 &  66 $\pm$ 30  & 200 $\pm$ 11\\
$i$ (mm) & $(^\circ)$       & 65 $\pm$  2  & [60]        & 66 $\pm$ 7   & 70 $\pm$ 3    & 44 $\pm$ 24 \\
$i$ (CO) & $(^\circ)$       & 62 $\pm$  5  & [60]        & 75 $\pm$ 10  & 70 $\pm$ 5    & 38 $\pm$ 11 \\
$V_\mathrm{LSR}$ & ($\kms$) & 6.42 $\pm$ 0.04 & 9.45 $\pm$ 0.28 & 5.88 $\pm$ 0.16 & 5.75 $\pm$ 0.15 & 6.8 $\pm$ 0.2 \\
$R_\mathrm{out}$ & (AU)     & 210 $\pm$ 7  & 110 $\pm$ 45 & 60 $\pm 20$ & 180 $\pm$ 24 & $> 150$  \\
$V_{100}$  & ($\kms$)       & 2.45 $\pm$ 0.29 & 1.8 $\pm$ 0.2 & 1.0 $\pm$ 0.1  & 2.45 $\pm$ 0.22 & 3.3 $\pm$ 0.3 \\
Log$_{10}(\Sigma_{100}(^{13}\mathrm{CO}))$ & (cm$^{-2}$)  &  16.5  $\pm$  0.5 & 15.2 $\pm 0.2$ &  16.0 $\pm$ 0.7 & 15.3 $\pm$ 0.1 &  15.2 $\pm$ 0.2 \\
Log$_{10}(\Sigma_{100}(\mathrm{C}^{18}\mathrm{O}))$ ($\diamond$) & (cm$^{-2}$)  &  15.0  $\pm$  0.1 &  & & 14.4 $\pm$ 0.2 & \\

\hline
\end{tabular}
\tablefoot{
($\diamond$) Using other parameters from $^{13}$CO.
($\dagger$) Numbers in bracket are values of fixed parameters.
(*) For information only: the fit misses substantial emission.}
\end{table*}

\paragraph{CW Tau}
\label{sec:sub:cwtau}
We have a clear detection of $^{13}$CO J=2-1 in CW Tau. \citet{Guilloteau+etal_2013} showed that
the line of sight towards this source is heavily affected by confusion with the surrounding
molecular cloud. Accordingly, channels near the systemic velocity are unreliable tracers
of the disk: confusion normally filters out emission, making the source appear smaller in these channels.
It would thus lead to \textit{higher} apparent inclination if those channels are included in the fit.
To avoid this bias, we ignored channels in the velocity range 5.29 to 7.62 km.s$^{-1}$, which result
 in an increase of the  the errors on systemic velocity and inclination. The rotation is consistent
with Keplerian rotation  (velocity exponent $0.50 \pm 0.04$) around a $ 0.64 \pm 0.08 \Msun$ star.
Because we ignored the central channels, the outer radius should be considered as a lower limit.

The errorbar on the $^{13}$CO column density appears unusually large compared to the other
sources: this is because the line is partially optically thick. However, the detection of
C$^{18}$O, with an integrated line flux around 0.2 Jy\,$\kms$, rules out large opacities
for $^{13}$CO, and can be used to obtain a better estimate. Assuming a standard isotopologue ratio of 8 between
$^{13}$CO and C$^{18}$O, the predicted column density of $^{13}$CO is $10^{15.7 \pm 0.1}$\,cm$^{-2}$,
similar to that of DE Tau.

The fit of the dust emissivity index $\beta$ is solely based on the 1.35 and 3.57 mm data, because
the continuum flux densities reported by \citet{Andrews+Williams_2005} appeared odd. Our
predicted flux at 0.85\,mm is 159 mJy, in excellent agreement with the result quoted in \citet{Andrews+etal_2013}.

CW Tau is of spectral type K3, and the  stellar mass derived by \citet{Bertout+etal_2007} from evolutionary tracks is $1.11 \pm 0.15 \Msun$, which is highly discrepant with our kinematical determination. \citet{Kenyon+Hartmann_1995} indicate an age estimate of 11 Myr. \citet{Andrews+etal_2013} have revised upwards the stellar luminosity to $\sim 2.4 \Lsun$, which (for a K3 spectral type) makes the star substantially
younger ($\sim 2-5$ Myr), but also more massive. The source would be much younger if one adopts a later spectral type in agreement with the kinematical mass, and would become more consistent with stellar ages in Taurus.

\paragraph{CX Tau}
In continuum, the apparent size is determined only
by the longest baseline data, and thus affected by phase noise.
After self-calibration, the data is consistent with a very marginally resolved disk.
However, the inclination is highly uncertain: any value above 20$^\circ$ is
acceptable.

There is a $5-6 \sigma$ detection of $^{13}$CO emission from this object (see Fig.\ref{fig:13co-cxtau}),
with an integrated line flux around 0.4 Jy\,km\,s$^{-1}$.
Using a fixed inclination of $\sim 60^\circ$ inclination, and assuming Keplerian
rotation, a disk model fit indicates a moderately resolved disk orbiting a star of
mass around 0.4 $\Msun$, in reasonable agreement with its
spectral type and derived stellar mass from evolutionary tracks (Table \ref{tab:prop}).

The systemic velocity, $V_\mathrm{LSR} \sim 9$\,km\,s$^{-1}$, appears unusual for the Taurus region,
where the stellar velocities cluster around 6\,km\,s$^{-1}$. However, the Taurus region has several
velocity components, and the molecular cloud velocity is around
8\,km\,s$^{-1}$ in this direction \citep[see Fig.12 of][]{Goldsmith+etal_2008}.
CX Tau is one of the objects classified as ``transition disks'' from its SED, because of the deficit
of NIR emission \citep{Najita+etal_2007}, which indicates a reduced opacity in the inner AUs in the IR domain.

\paragraph{DE Tau}
There is a weak $^{13}$CO detection, with an integrated line flux around 0.4 Jy\,km\,s$^{-1}$.
Orientation and inclinations agree with those derived from the continuum emission, with large uncertainties
given the small disk size, and the derived dynamical mass is $\sim 0.11 \Msun$, but the
fit is rather poor (see Fig.\ref{fig:13co-detau}).
DE Tau is of spectral type M2, which would be more compatible with a stellar mass of $0.4 \Msun$
(see Table \ref{tab:prop}).

\paragraph{DS Tau}
This source has the largest dust disk in our sample, with a characteristic radius twice larger
than any other one. The continuum visibilities suggest the existence of an inner hole. The best fit radius
for such a hole is $15 \pm 5$ AU in the power law model, and slightly larger, $20 \pm 4$ AU,
for the viscous profile.

There is a clear $^{13}$CO detection, with an integrated line flux
around 0.5 Jy\,km\,s$^{-1}$
(see Fig.\ref{fig:13co-dstau}).
The disk parameters are well determined, and indicate a small gas disk (180 AU outer radius), with
low $^{13}$CO surface density. The rotation is consistent with a Keplerian law (exponent $0.50 \pm 0.08$), 
and the implied stellar mass is $0.77 \pm 0.07 \Msun$.
The  $^{13}$CO data are also consistent with the existence of
an inner cavity, with a best fit inner radius of $30 \pm 9$ AU. C$^{18}$O is at the limit of detection,
with an [$^{13}$CO/C$^{18}$O] isotopologue ratio of about $\sim 8$.

DS Tau is part of a visual binary, with a 1 magnitude fainter companion (called DS Tau/c or DS Tau B) 
located $\sim 7.1''$ to the North-West \citep[at PA 294$^\circ$,][]{Moneti+Zinnecker_1991}, but DS Tau 
is usually considered as a single star, because the companion has no H$\alpha$ emission and no IR excess 
\citep{Hartmann+etal_2005}. Both \citet{Moneti+Zinnecker_1991} and \citet{Leinert+etal_1993} conclude 
that the probability of chance association with a background star is high, especially as the cloud 
visual extinction is low in this direction.  However, we have  a $\sim 4.5 \sigma$ (0.7 mJy) signal 
exactly at this position (7.2$''$ at PA 292$^\circ$, see Fig.\ref{fig:large}), which argues that 
the visual binary is indeed a common proper motion pair comprising a cTTs and a wTTs/Class III source.


\paragraph{FM Tau}
This source 
is the most impacted by self-calibration. The analysis favors a marginally resolved, highly inclined 
disk but the errorbars on the inclination are large: any value
above 20$^\circ$ is acceptable at the $3 \sigma$ level.

\paragraph{FZ Tau}
FZ Tau 
also appears very compact. The inclination
is $< 70^\circ$ at the $3 \sigma$ level. The position angle is only constrained if the inclination
is $> 40^\circ$.

\paragraph{HO Tau}
The disk is well resolved, with a similar size to that of CW Tau. However, no $^{13}$CO is detected.
Assuming the gas disk would be of similar size to those of CW Tau or DS Tau, the non
detection implies an upper limit of $10^{14.8}$\,cm$^{-2}$ for the $^{13}$CO surface density at 100 AU, i.e.
4 to 10 times smaller than in the other detected disks.

HO Tau appears to be the oldest star in our sample, even with the
age of 9 Myr indicated by \citet{Ricci+etal_2010a}.

\paragraph{SU Aur}
SU Aur was not properly centered in the primary beam(\footnote{A typo in \citet{Ricci+etal_2010a} 
give a position off by 9$''$ in declination}) in the initial observations. This pointing error 
was corrected for the A configuration,
and a correction for primary beam attenuation was applied to the mispointed data sets  (a factor 
$\sim 1.7$ given the $9''$ offset).

The continuum is essentially unresolved. This can be consistent with
a 10 AU radius optically thick disk with $T_{100} = 15$ K. However, the high submm flux
\citep[$\sim 70$ mJy at 850 $\mu$m][]{Andrews+Williams_2005}.
favors a warmer disk, which is then somewhat optically thinner. With $T_{100} = 30$ K, a better global
solution is found. The size may also be slightly underestimated because for the most compact configurations,
the source was placed near the half power beam, and thus the amplitude calibration is a critical
function of pointing errors. However, the amplitude error should not exceed 20\%, still leaving the
source quite marginally resolved.
The quoted sizes and $\beta$ assume $T_{100} = 30$ K.

$^{13}$CO is clearly detected.  The total flux is $\sim 1.1 \pm 0.4$ Jy\,km\,s$^{-1}$.
The derived line parameters are in good agreement with the 30-m data of \citet{Guilloteau+etal_2013}
(line flux $1.7\pm0.4$ Jy\,km\,s$^{-1}$, systemic  velocity $7.1\pm 0.3$ km\,s$^{-1}$
line width $3.0 \pm 0.5$ km\,s$^{-1}$). The $^{13}$CO emission is rather complex.
There is extended emission, spread about $\sim 3-5''$ in the East-West direction,
with a relatively narrow line, and the bulk of the emission is offset by $\sim 0.5''$ from the continuum
peak (see Fig.\ref{fig:13co-suaur}). It is possible to model the observed
emission by a Keplerian disk orbiting a $\sim 1.2 \Msun$ star, centered on the continuum emission peak,
However, such a model misses emission North of the star, and the disk parameters are not well constrained,
apart from the $^{13}$CO column density.

The disk inclination and orientation derived from CO emission agree well with the inner disk rim 
properties inferred from IR interferometric studies by \citet{Akeson+etal_2002}. An IR nebulosity 
was detected by \citet{Chakraborty+Ge_2004} South-West of the star, while an optical emission 
extending few arcsec westwards of the star has been reported by \citet{Nakajima+Golimowski_1995}. 
Our disk orientation favors the first scenario proposed by \citet{Chakraborty+Ge_2004}, where the 
IR nebulosity would result from the opening of a cavity in the surrounding molecular cloud by 
the stellar outflow. We note that degrading the spatial resolution in the $uv$ plane results 
in a detection of extended $^{13}$CO emission spreading a few arcsec from the star in the western 
direction, with a good overlap with the optical emission seen in HST images. In any case, the 
IR nebulosity extends much farther out than the disk outer radius reported in our study (even 
in $^{13}$CO) and is more likely to trace a residual envelope or parent cloud emission rather 
than the disk itself.

\paragraph{V 836 Tau}
is one of the few objects with characteristics at the limit between cTTs and wTTs from its 
variable H$\alpha$ emission, and, from its SED, between Class II and III \citep{Najita+etal_2008}. 
As such, it may be a system in which the disk is being dissipated and just becoming optically
thin. We find that the dust emission is largely resolved, with a characteristic radius $R_c \sim 25$ AU. 
It is similar in many respect to HO Tau. Here, the low molecular content is attested by its very weak 
$^{12}$CO emission \citep{Duvert+etal_2000}. The $\beta$ index determination uses the 3\,mm flux 
densities from \citet{Duvert+etal_2000}, $3.0 \pm 0.8$ at 3.4 mm and $7.7 \pm 1.2$ at 2.7 mm, and the 
850 $\mu$m data from \citet{Andrews+Williams_2005}.

\begin{figure*}
  \sidecaption
   \includegraphics[width=14.0cm]{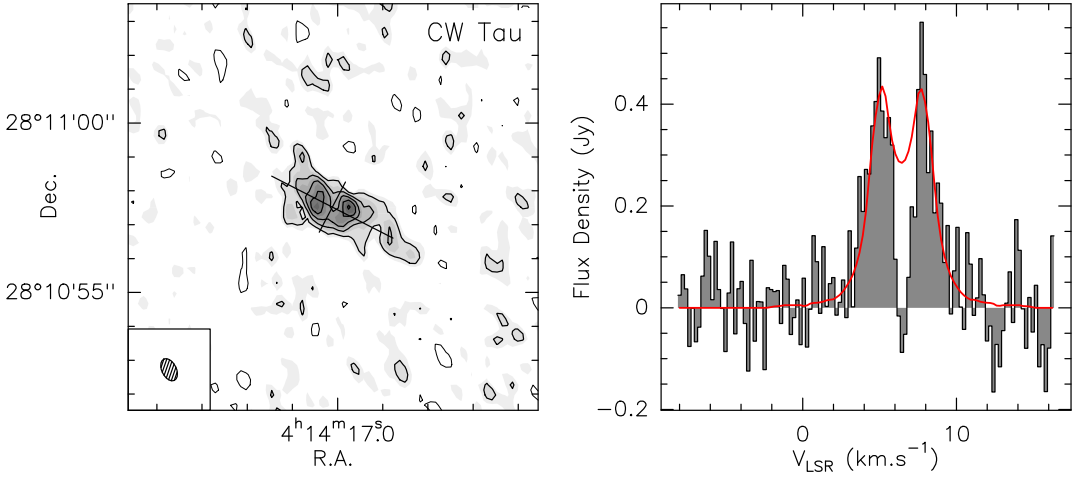}
 \caption{Left: S/N map of the $^{13}$CO 2-1 emission towards CW Tau.
 The cross indicates the position, orientation and aspect ratio
 of the continuum disk. Contours are in steps of 2 $\sigma$, with the
 zero contour omitted and negative contours dashed. Right: Integrated $^{13}$CO 2-1 spectrum.}
  \label{fig:13co-cwtau}
\end{figure*}

\begin{figure*}
  \sidecaption
   \includegraphics[width=14.0cm]{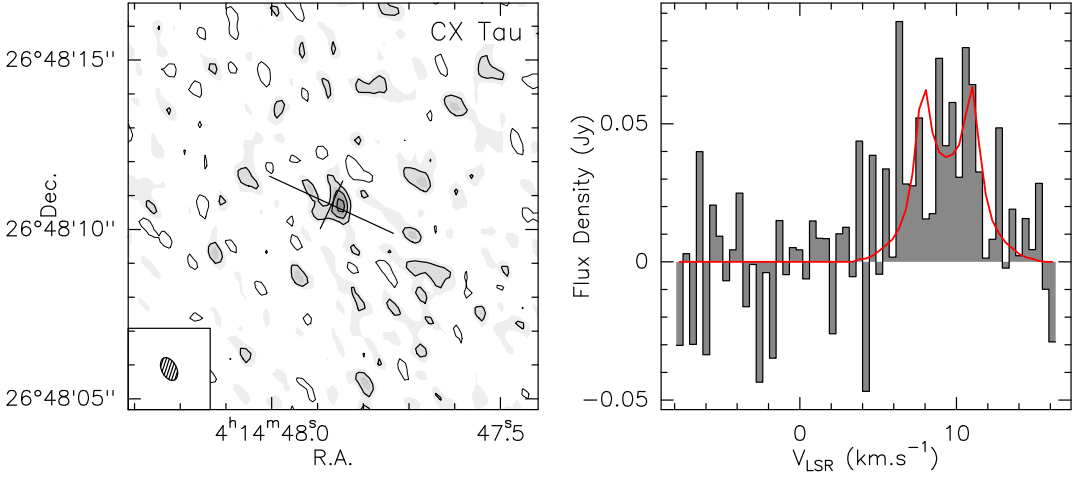}
 \caption{As Fig.\ref{fig:13co-cwtau} but for CX Tau.}
  \label{fig:13co-cxtau}
\end{figure*}

\begin{figure*}
  \sidecaption
   \includegraphics[width=14.0cm]{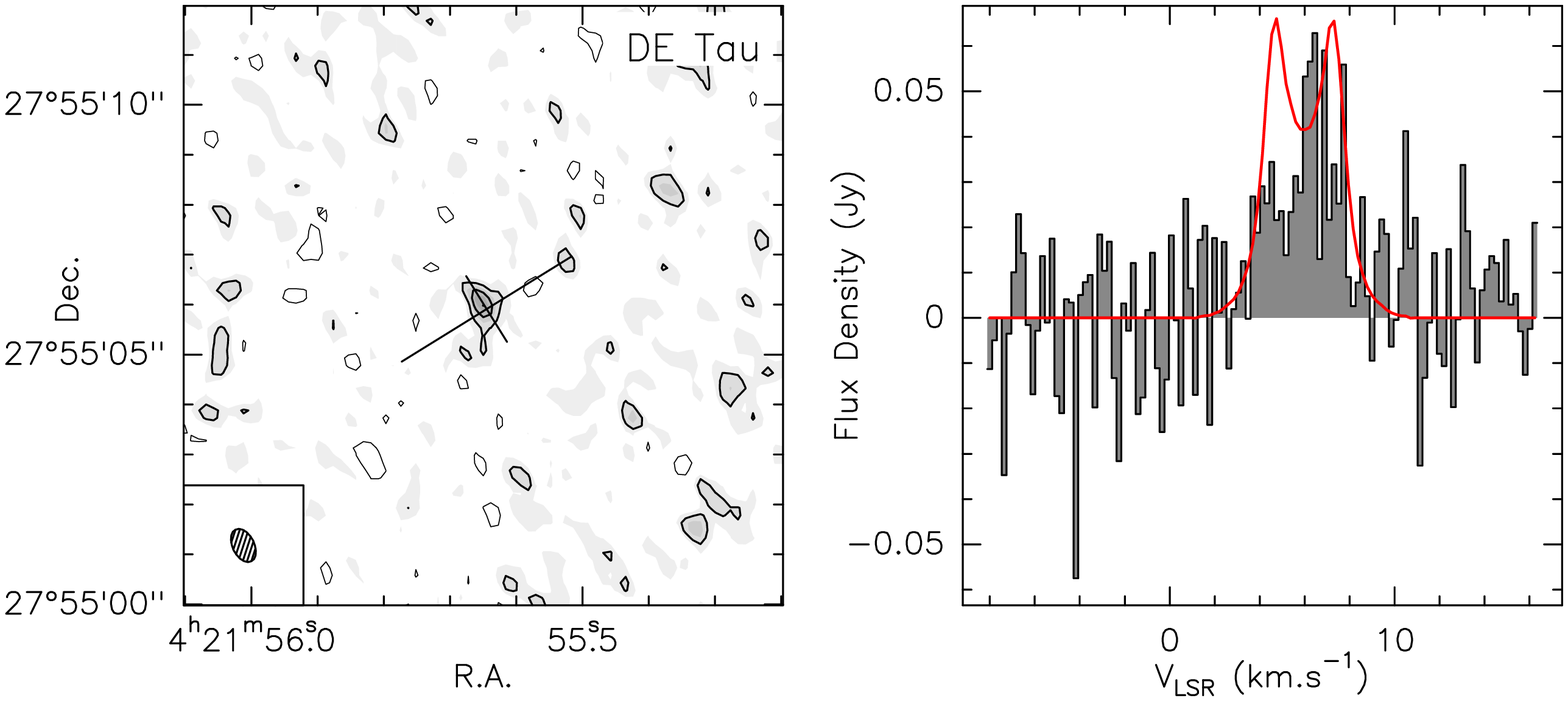}
 \caption{As Fig.\ref{fig:13co-cwtau} but for DE Tau.}
  \label{fig:13co-detau}
\end{figure*}
\begin{figure*}
  \sidecaption
   \includegraphics[width=14.0cm]{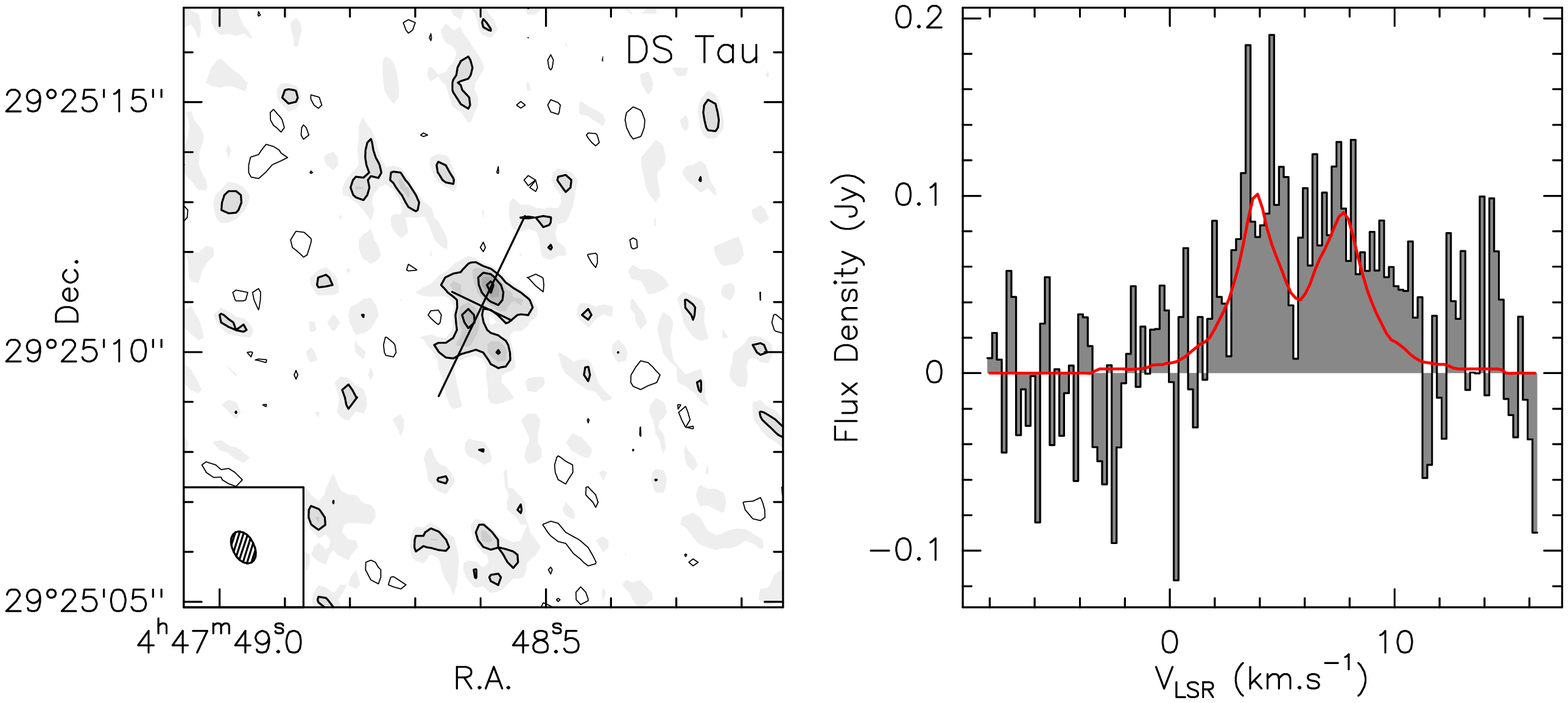}
 \caption{As Fig.\ref{fig:13co-cwtau} but for DS Tau.}
  \label{fig:13co-dstau}
\end{figure*}

\begin{figure*}
  \sidecaption
   \includegraphics[width=14.0cm]{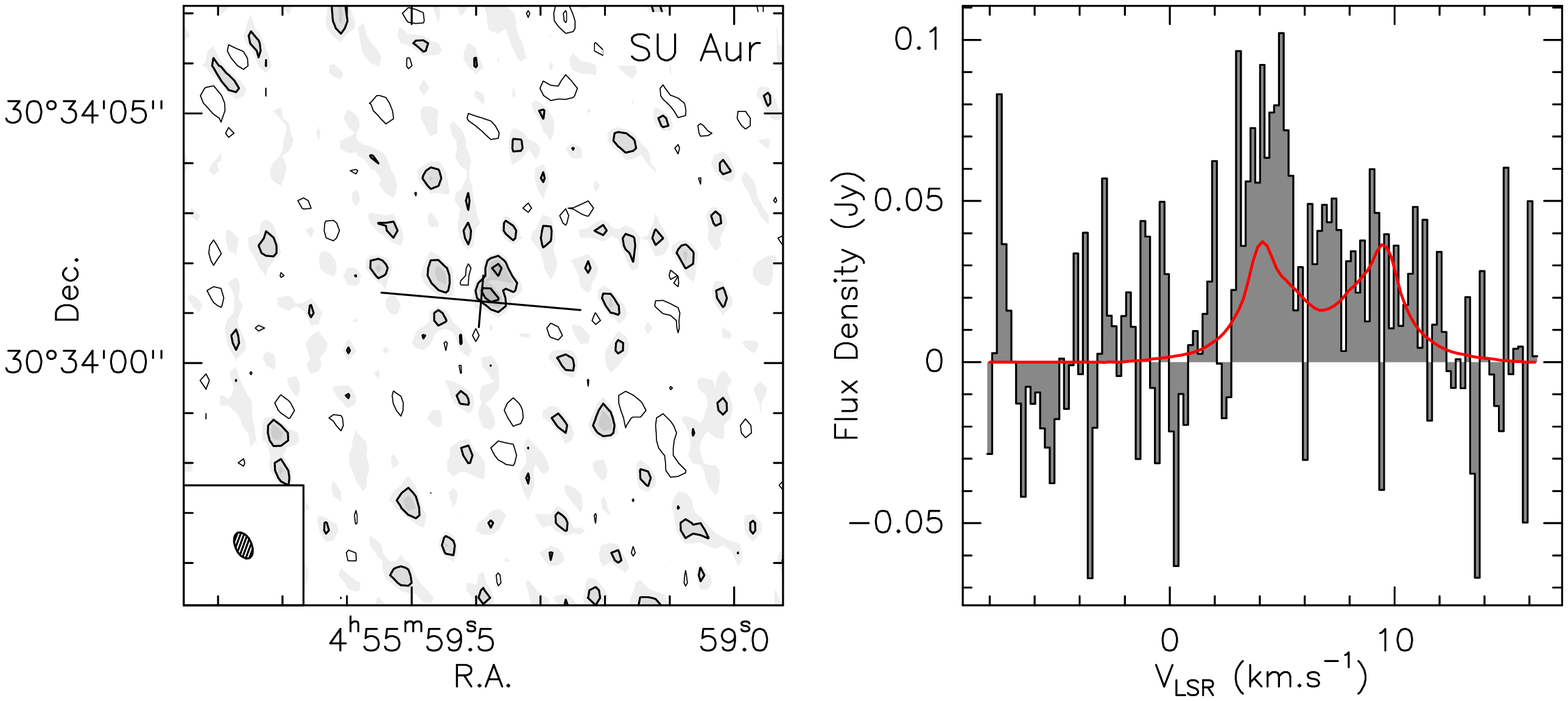}
 \caption{As Fig.\ref{fig:13co-cwtau} but for SU Aur.}
  \label{fig:13co-suaur}
\end{figure*}

\section{Discussion}

\subsection{General properties}

From Table \ref{tab:self}, half of the sources in our sample are clearly very compact: CX Tau, FM Tau, FZ Tau and 
SU Aur  are at best only marginally resolved in continuum. For these sources, in practice, we can only set upper 
limits on $R_c$ and, in favorable cases, lower limits on $i$.
In these objects, the density profile is not
constrained by our observations, and optically thick disks would represent the observations equally
well. With our adopted temperature profile, a (face on) optically thick core of outer radius $R_\mathrm{thick}$
has a flux of
\begin{equation}\label{eq:fluxthick}
    S_\nu  = \frac{2 k T_0}{D^2} \frac{\nu^2}{c^2}  \frac{2 \pi R_0^2}{2-q}
    \left( \left(\frac{R_\mathrm{thick}}{R_0}\right)^{2-q} - \left(\frac{R_\mathrm{int}}{R_0}\right)^{2-q} \right)
\end{equation}
which for $T_0=15$ K, $R_0=100$ AU and $q=0.4$ is 15 mJy at 220 GHz for $R_\mathrm{thick} = 7$ AU at
the Taurus distance (D=140 pc). Although the errorbars of the characteristic radius are difficult to ascertain for the most
compact sources, a simple look at the visibility profile indicate that these sources must be smaller than
the smallest resolved one, DE Tau, implying $R_\mathrm{out} < 20$ AU, but still large enough to provide
sufficient flux, implying $R_\mathrm{out} > 7$ AU. A more complete exploration of the error bars is
given in Appendix \ref{app:errors}. As expected, the smallest disks, which are essentially unresolved and
have flux density in the range 9 - 15 mJy, are consistent with being optically thick.
Our adopted opacity being $\kappa(220 \mathrm{GHz}) = 0.02$ cm$^{2}$g$^{-1}$, this implies a minimal surface
density of 50 g\,cm$^{-2}$ and a minimal disk mass of $\sim 10^{-3} \Msun$ for such disks.
Given the errors, we actually find only minimal surface densities of $7$ g\,cm$^{-2}$ for FM Tau and CX Tau,
and $20$ g\,cm$^{-2}$ for FZ Tau and DE Tau.
Higher temperatures would reduce the required outer radius as $1/\sqrt(T)$ and mass as $1/T$ (this temperature
dependence explains while SU Aur is optically thin).

The small sizes are also confirmed by the CO isotopologue results. The gas disks also are
much smaller than the $\sim 500$ AU more typical of brighter disks. However,
as in most other resolved disks, the spatial extent
in the CO isotopologues appears to be substantially larger (by a factor 2 to 6) than the characteristic radius
in the continuum.

Because the majority of the disks are essentially unresolved, we also cannot constrain the dust emissivity
index $\beta$, since the whole disk can be optically thick at all wavelengths shorter than 3\,mm. We thus have
significant measurements of $\beta$ only in the 4 disks which are large
enough to rule out significant optical depth: the brightest one, CW Tau, and 3 faint ones, HO Tau, DS Tau and V836 Tau.

\subsection{Origin of disk faintness}

Although the sample is limited in size, a number of properties emerge from
this study.

\paragraph{Small Disks}
Small disk size is the main cause of low disk flux. A substantial fraction of these ``faint'' disks, of
order 50 \% in our sample, are so small that they remain essentially unresolved,
with characteristic sizes $\leq 10$ AU. These
small disks are essentially consistent with being optically thick down to
3\,mm, a result which by itself can explain the apparent $\alpha = 2$.
The minimal surface density at $\sim 10$ AU is on the order of
50 g\,cm$^{-2}$: this is quite comparable with what is derived for much larger
and brighter disks like CY Tau, DM Tau or MWC 480 \citep[see][their Fig.12]{Guilloteau+etal_2011}.
Evidence for grain growth in these very small disks is lacking, as no observation reaches
the optically thin regime so far.

\paragraph{Inner Cavities}
We have a tentative detection of an inner cavity in DS Tau. The missing flux which might be attributed to
the existence of the cavity is not large enough to explain alone the low mm flux of DS Tau: its
contribution is would be at most 10 mJy for DS Tau. As CX Tau has been classified as a ``transition disk'' from its near-IR SED, the DS Tau cavity makes the fraction
of ``transition disks'' potentially as high as 20 \% in our sample, comparable to what has been found
for brighter, larger disks \citep[e.g.][]{Andrews+etal_2011}. \citet{Najita+etal_2007} argued that this fraction
was decreasing with disk mass (or more precisely, continuum flux). \citet{Owen+Clarke_2012} suggested
that ``transition disks'' are made of two distinct populations. However, the identification of central
cavities has often been based on IR modeling only. In practice, cavities of 20-40 AU radii
were often detected at mm wavelengths: e.g. the first detection in LkCa 15 by \citet{Pietu+etal_2006},
or HH 30 \citep{Guilloteau+etal_2008}. Confirming the existence of cavities in fainter sources is
a challenging task which require higher angular resolution and sensitivity only reachable by ALMA.

\paragraph{Low surface density:}
We have two well resolved disks with similar sizes (CW Tau and HO Tau, $R_c \sim 20$ AU), but
widely different $\beta$. On one hand, the fainter one has $\beta \simeq 0$, so that its lower flux may be
explained  by grain growth rather than by a lower disk mass / surface density. On the other hand,
$^{13}$CO is only detected in the brighter disk: this suggests that the
gas content of HO Tau is indeed smaller than that of CW Tau. V836 Tau, a transition
object between Class II and Class III types, appears very similar to HO Tau, although
with larger errors.

For the other objects, the $^{13}$CO surface densities at 100 AU are comparable to those
measured at 300 AU in the large disks studied so far such as \object{DM Tau}, \object{LkCa15} 
or \object{MWC 480} \citep{Pietu+etal_2007}. Extrapolation of the later values to smaller radii is difficult because
of the substantial optical depth in $^{13}$CO: \citet{Pietu+etal_2007} find rather large $p$ exponents, but
a flattening of the surface density distribution inwards is not excluded. The newly detected disks only
appear as down-sized versions of the large ones, but do not necessarily have lower surface densities 
in the inner 10 to 20 AU.

\begin{figure}
   \includegraphics[width=\columnwidth]{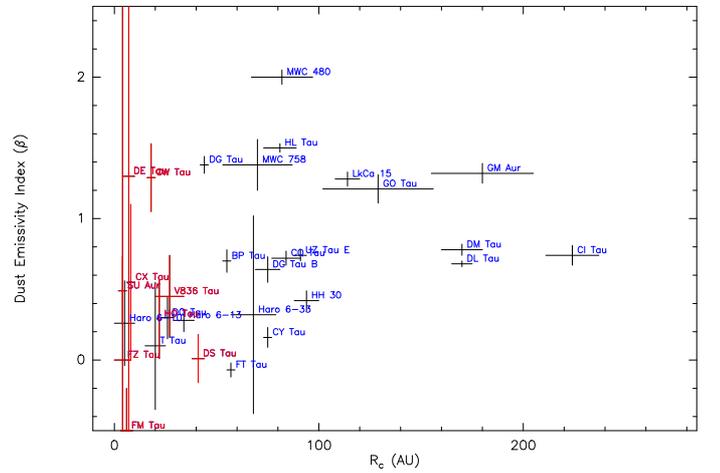}
 \caption{Dust emissivity index $\beta$ as a function of characteristic radius $R_c$.
 Sources in blue are from \citet{Guilloteau+etal_2011}, and in red from this work.}
  \label{fig:edges-beta}
\end{figure}

\paragraph{Grain Growth:}
Direct evidence for grain growth exists for 3 of the 4 resolved disks: DS Tau, HO Tau
and V836 Tau, which all have $\beta \sim 0$. However, we cannot determine the relative
importance of grain growth in lowering the mm emission compared to a possible low
surface density, as these 3 sources have various degree of $^{(13)}$CO surface densities.
Because of the possible contribution by an optically thick core,
the dust emissivity index is ill constrained  in the more compact sources,
and the $\alpha-2$ value (where $\alpha$ is the spectral index of the emission)
is only a stringent lower limit.  In fact, there is no significant correlation
between the disk ``size'' $R_c$ and $\beta$, as shown by Fig.\ref{fig:edges-beta}.
Such a correlation is expected on the basis of enhanced dust growth in the
inner parts of the disk, which leads to a variation of the dust emissivity index
$\beta$ as a function of radius. Smaller values of $\beta$ for radii smaller than
$\approx 60$ AU has been found by \citet{Guilloteau+etal_2011} for
all sufficiently resolved disks in their sample, except of course those
with large inner cavities devoid of dust. A similar result has been obtained for AS 209 with
higher signal to noise by \citet{Perez+etal_2012} through multi-frequency observations
with sub-arcsec resolution.
Unfortunately, sources in the current sample are
insufficiently resolved out to extend this property to small disks.

CW Tau, which is small but has $\beta \sim 1.3$,  appears as an exception to the trend found
by \citet{Guilloteau+etal_2011}. Although its evolutionary status is unclear because of
the discrepancy between the derived dynamical mass and spectral type, it may be quite
young given its relatively high luminosity, so that its high $\beta$ value could be
related to this young age.

DS Tau, which exhibits a (relatively) large disk with low surface density, inner cavity,
evidence of grain growth, and forms a common proper motion pair with a Class III / wTT star, may be
a case of a system caught in the stage of dissipation. We note however that DS Tau still exhibits
a substantial accretion rate, much larger than expected for photo-evaporation to play a significant
role in the disk dissipation \citep[e.g.][]{Gorti+etal_2009}.

\subsection{Towards a new population of small, dense disks ?}

\begin{figure}
   \includegraphics[width=\columnwidth]{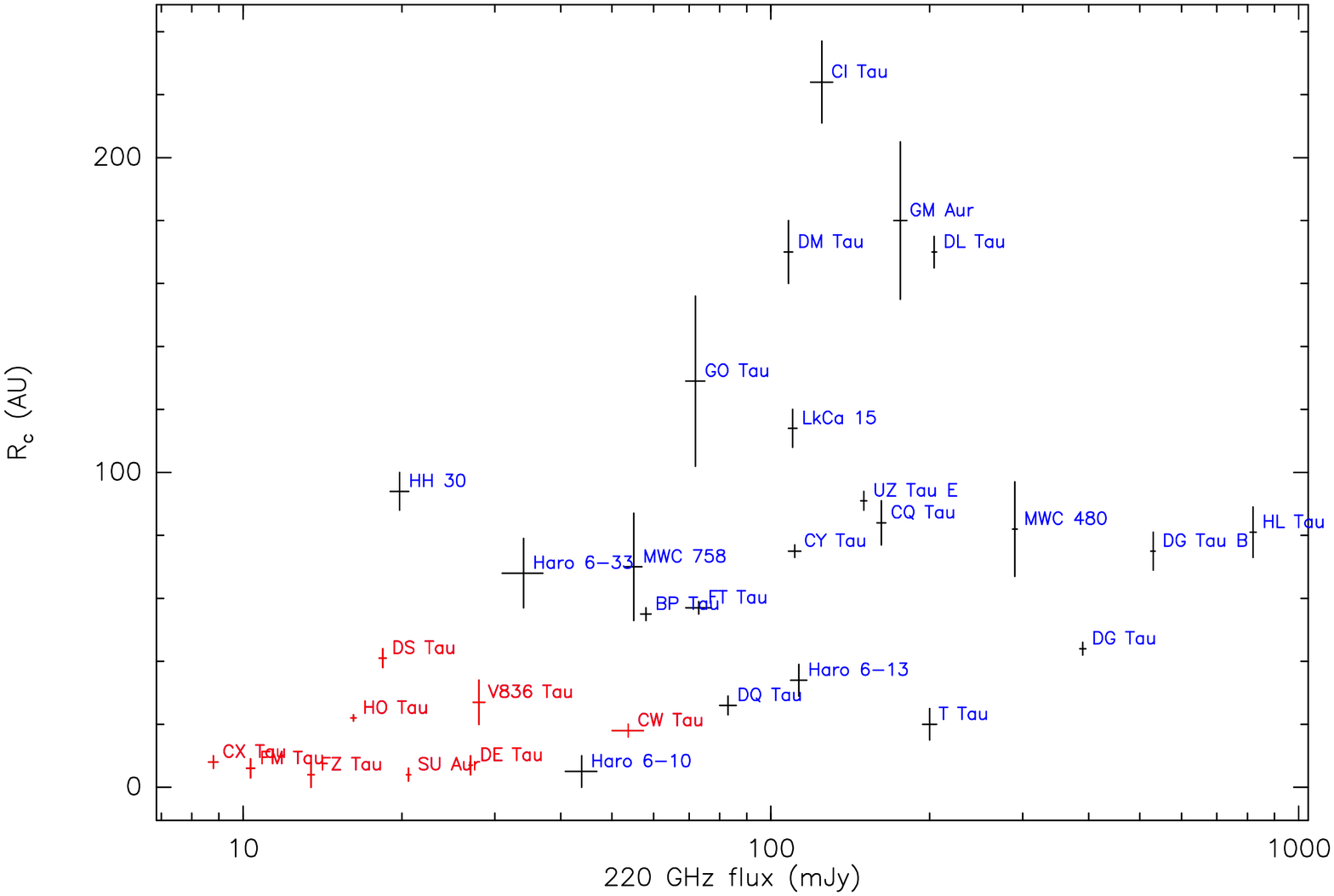}
 \caption{Characteristic radius  $R_c$ as a function of 230 GHz flux density.
 Sources in blue are from \citet{Guilloteau+etal_2011}, and in red from this work.}
  \label{fig:edge-flux}
\end{figure}

Although the observed disks show a range of sizes, they are all quite substantially
smaller than those studied by \citet{Guilloteau+etal_2011}. This is clearly
demonstrated by Fig.\ref{fig:edge-flux} which shows
a scatter plot of the characteristic radius $R_c$ \citep[re-derived from the results of][under the
assumption of $\gamma=0.5$]{Guilloteau+etal_2011} and 230 GHz flux density.

Not only the dust disks, but also the gas disks appear significantly smaller
than in previously studied sources. The gas disk radii range between 50 and 220 AU, while
the 30 disks with molecular emission detected by \citet{Guilloteau+etal_2013} (out of a sample of 47 objects) had outer
radii $> 300$ AU, the sensitivity limit of that study. Since the size distribution in
\citet{Guilloteau+etal_2013} peaked near 300 AU, it is thus conceivable that the faint
disks detected here constitute the low end tail of the disk size distribution.

\begin{figure}
   \includegraphics[width=0.8\columnwidth]{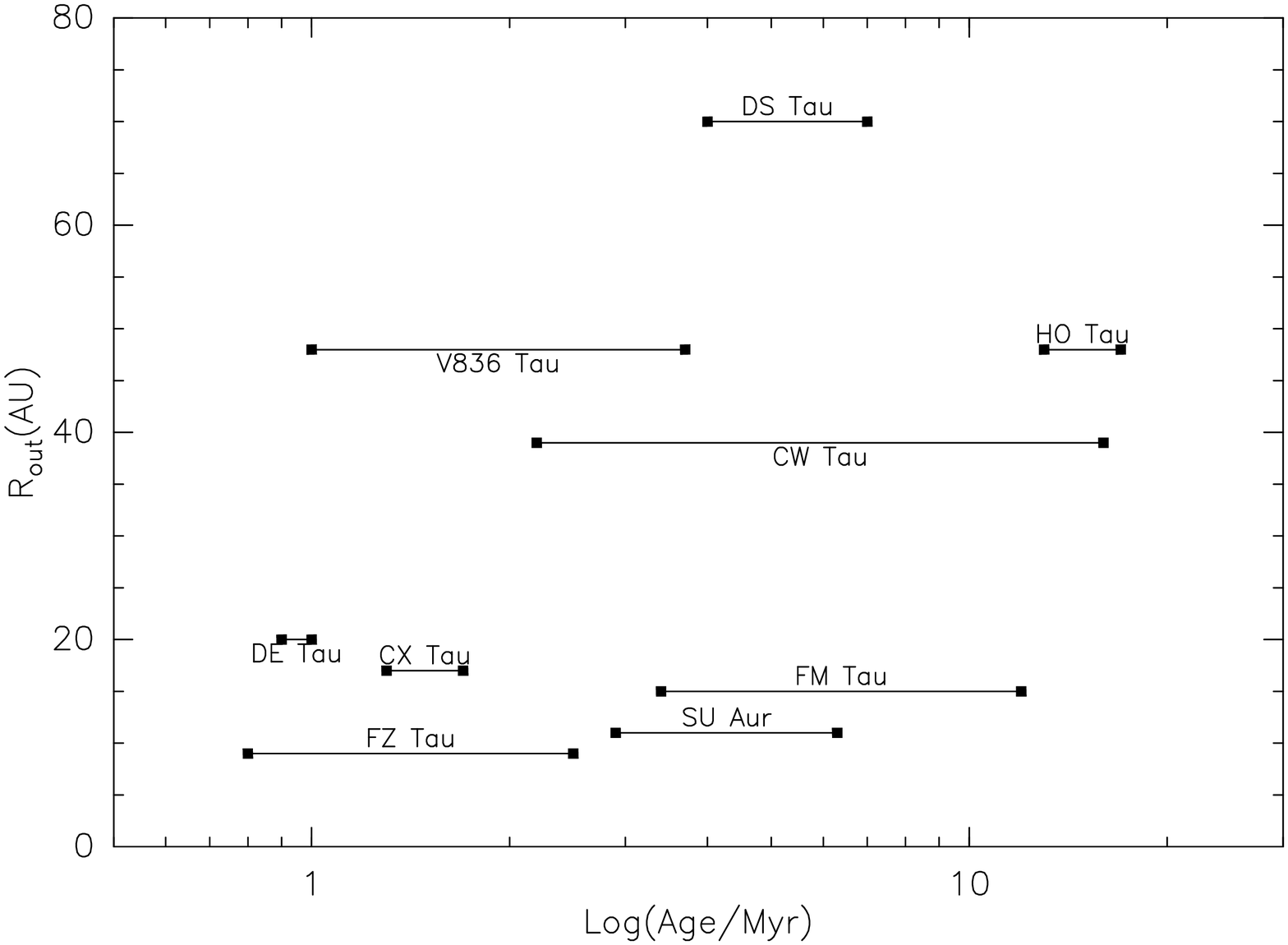}
 \caption{Dust disk radius $R_\mathrm{out}$ as a function of the system Age.}
  \label{fig:ages-radius}
\end{figure}

\begin{figure}
   \includegraphics[width=0.8\columnwidth]{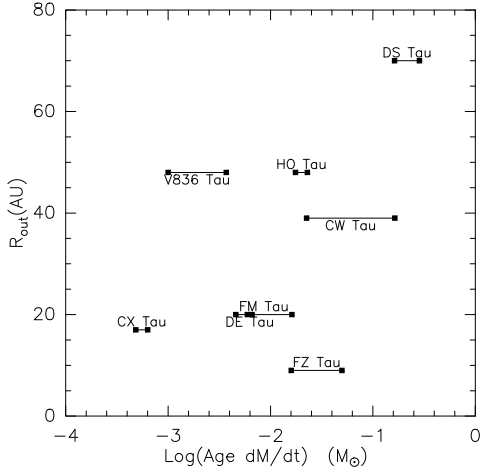}
 \caption{Dust disk radius $R_\mathrm{out}$ as a function of the product Age $\times$ Mass accretion
  rate.}
  \label{fig:accreted-radius}
\end{figure}

Our study reveal for the first time a population of very small (dust and gas) disks,
but whose surface density in the inner 10 AUs does not necessarily differ from that of the
much larger disks studied so far.  There is no obvious explanation for this difference
in sizes.  We briefly explore here some possibilities.
\begin{itemize}
\item \textit{Tidal effects:} \\
Tidal truncation does not appear to play a significant role, as all
sources, except DS Tau, are single.  Binary separation in the range
40 - 500 AU would be required to restrict the disk sizes (in line and continuum)
to the observed values, while no companion has been found at such distances with
typical contrast limits of about $\Delta m \simeq 5$ from IR/optical studies
of these stars \citep[e.g.][for even closer companions]{Ghez+etal_1993,Kraus+etal_2011}. For
the wide visual binary DS Tau, tidal effects must be small, unless the companion is on a
very eccentric orbit. Furthermore, DS Tau is the largest disk in our
sample.
Only one other source shows a possible emission from a hidden companion:
a 6 $\sigma$ signal is found 3.8$''$ NE of CX Tau, with a flux of 0.7 mJy,
but at this level, contamination by extragalactic high redshift sources is
not excluded. It is also unlikely that these small disks are due to
past tidal truncation in multiple systems, followed by ejection of the observed
component, as the stars velocities (Table \ref{tab:coline})
and proper motions  \citep{Ducourant+etal_2005} do not
deviate from those of the Taurus association members.
Only CX Tau exhibits a rather unusual velocity,
but it differs only by $\sim 1.5$\,km\,s$^{-1}$ from that of surrounding molecular cloud
(see Sec.\ref{sec:results}).
\item \textit{Age:}\\
Disks around the youngest objects (e.g. HL Tau, DG Tau, Haro 6-13) were found to
 be more centrally peaked
 by \citet[][see their Fig.12]{Guilloteau+etal_2011},
 possibly because they have not yet been spread out by viscous dissipation. Here,
 the observed stars are not particularly young (see Fig.\ref{fig:ages-radius}). Some
 even have estimated ages which are at the high end of ages in the Taurus
 Auriga region, in excess of 10 Myr. Note that ages may not all be trustable
 (specially for CW Tau, see Sec\ref{sec:sub:cwtau}).
\item \textit{Low viscosity}:\\
Lower viscosity may have kept the disks small. Indeed, these faint disks
have somewhat lower accretion rates than average \citep[taking the median value
log$(\mdot) = -7.6$ from][]{Najita+etal_2007}. However, there is no clear correlation between
the product of the mass accretion rate $\dot{M}$ with the age of the star $t_*$,  $(\dot{M} t_*)$ 
and the characteristic radius $R_\mathrm{out}$, although this would
be expected in a viscous evolution model since both quantities scale as the effective
viscosity (see Fig.\ref{fig:accreted-radius}).
Moreover, we have to explain here two
very different sizes: the gas disk sizes,
$\sim 100-200$ AU, which can easily be accommodated by somewhat reduced viscous
spreading, and the dust disk sizes, which are much smaller ($\sim 10 - 30$ AU).
\item \textit{Layered Dead Zone:}\\
 one possibility in this direction would
be that the observed dense, possibly optically thick, continuum cores
are direct manifestations of the (magnetically) dead zones, as predicted by models of \citet{Zhu+etal_2010},
with $^{13}$CO line emission coming from the more viscous and diffuse surrounding disk. 
Our sensitivity is insufficient to detect line emission from the inner 20 AU, making us
unable to probe the change in surface density predicted by these models.
Note that brighter, larger, disks may also exhibit optically thick (continuum) cores of size comparable
to the ones found in our sample, and may only differ from the fainter ones by the \textit{outer,
viscous} disk component.
\item \textit{Planetary System Formation:}\\
We may speculate that the very compact dust disks are due to dust
shepherding by \protect{(proto\-)planets} at a few AU
from the stars, which would result in a dense ring of large grains just outside the
planet's orbit.  Gas and small dust would be much less affected, and could still extend
out to 150 - 200 AU by viscous spreading.  CX Tau, which displays signs of an evacuated
inner cavity through its NIR deficit, and DS Tau, where our data reveal a potential
$\sim 20$ AU radius cavity, are good candidates to probe such a mechanism. Much
higher angular resolution studies will be needed to explore this possibility.

\item \textit{Just born small ?}\\
Perhaps the simplest alternative is that such disks are just born small because they originate from
the low-end tail of the initial angular momentum distribution of collapsing cores.
A quantitative comparison is however difficult, because of large uncertainties on
both sides. For the disks, the extrapolations of the initial radii
are extremely hazardous \citep[see][]{Guilloteau+etal_2011}, and the total angular
momentum is sensitive to the surface density in the outer parts, which is ill-constrained. On the
other hand, for molecular clouds, the derivation of angular momentum from observations remains difficult \citep{Dib+etal_2010}.
\end{itemize}

\subsection{Consequences}

\label{sec:tdust}
\citet{Andrews+etal_2013} found that 50 \% of T Tauri stars in the Taurus Auriga region
have mm flux $< 25$ mJy at 1.3\,mm.  Our limited sample suggests that half of them could be due
to very compact disks with high surface densities.  The other half is a combination of
several effects. Grain growth obviously plays a role, but the $\beta$ values found for these disks
is not different from that found in the inner parts of big disks by  \citet{Guilloteau+etal_2011}.
Disk dispersal (a reduction in surface density) is difficult to disentangle from grain growth, although
the low $^{13}$CO line flux suggests that it may affect 30 to 40 \% of the sources.  Note that
these conclusions are only valid for stars of mass $> 0.3 - 0.4 \Msun$, as our sample include
no stars with spectral type later than about M2.

The existence of a category of very small disks has consequences on the
relation between disk mass $M_d$ and mm flux $S_\nu$:
\begin{equation}
\log(M_d) = \Delta + \log(S_\nu) + 2 \log(D) - \log(\kappa_\nu) - \log(B_\nu(T_d)) .
\end{equation}
where $D$ is the distance, $T_d$ the effective mean dust temperature, and $\Delta \ge 0$
an opacity correction factor. $\Delta = 0$ for optically thin dust, but can become
arbitrary large if the optical depth is significant. While $\Delta$ can be shown
to be in general small for large disks \citep[radii $> 100$ AU, e.g.][]{Beckwith+Sargent_1991,Andrews+etal_2013},
it may be dominant for the very compact disks.

The small disk size also has an impact on the mean temperature, as the dust temperature
decreases with distance to the star. For $R_\mathrm{out} = 20$\,AU, $T_0 = 15$\,K
at $R_0 = 100$ AU and $q = 0.4$, we find $T_d = 35$\,K, a value which is somewhat
above the $T_d = 25 (L_*/\Lsun)^{1/4}$\,K found by
\citet{Andrews+etal_2013} using a 200 AU radius disk model.
The combination of optical depth and temperature effects will
influence the dispersion around the $S_\nu / M_*$ correlation. These effects must be kept
in mind when using flux densities as a proxy for disk masses, as such conversion
often implicitely assume that disks are sufficiently large.

\citet{Birnstiel+etal_2010} used a comprehensive dust disk evolution model including
coagulation and fragmentation and compared its predictions to the Taurus/Ophiucus observations of \citet{Ricci+etal_2010a}.
While they could reproduce the spectral index distribution between 1 and 3\,mm, the model was overpredicting the 1\,mm flux.
However, they only considered a single disk surface density shape with $R_c=60$ AU and $\gamma=$1. Our results indicate
that a substantial fraction of the fainter disks have much smaller sizes, and that this is the main reason for
the weaker fluxes.

The substantial surface densities in the inner 10 AUs also implies that giant planet
formation can still occur in these overall less massive, but much smaller, disks.  The existence
of a population of very compact disks may be related to the closely packed compact planetary systems discovered
by the Kepler mission, like
Kepler 11 \citep{Lissauer+etal_2013} or Kepler 33 \citep{Lissauer+etal_2012},
since the inner regions of most faint disks seem to retain enough gas content to form multiple systems of low-mass gaseous planets in the first AUs.

Finally, it may be worth stressing that essentially all estimates of surface densities
in the inner 10 AU for large disks are extrapolations from larger radii. Apart from
disks with inner cavities, current observations cannot rule out that most disks exhibit
a dense inner core of $\sim 10$ AU radius which remains optically thick at all mm wavelengths.

\section{Conclusions}

We report the first high angular resolution study of disks with low level of continuum
emission at mm wavelengths. We find that these disks are substantially smaller than
the brighter ones imaged so far. A majority of them remain essentially unresolved at 0.4$''$
resolution, leading to radii $< 15$ AU. These disks are small enough to be optically
thick at 1.4 mm, which provides a simple explanation for their apparent spectral
index $\alpha \sim 2$.

$^{13}$CO emission has been detected in five disks. Although the gas distribution extends
 further out than the apparent size of the dust emission (as in all other disks around
 T Tauri and HAe stars), its outer radius is also small compared to previously studied
 gas disks.

Only two disks (V836 Tau, an intermediate case between Class II and Class III sources, and HO Tau)
have apparently low surface densities simultaneously attested by their low opacity in continuum
and by their low molecular content. These stars may be caught in the act of dissipating their disks.
Despite its higher $^{13}$CO content, DS Tau, which forms a common proper motion pair with
a Class III/wTT star, is another good candidate for a dissipating disk because of the
possible existence of an inner cavity.

This study reveals a population of small disks whose surface density remains large
in the central 10 AU region. This population had been ignored so far because
of the limited sensitivity and angular resolution of previous observations. It may
represent up to 25 \% of the whole disk population in Taurus. These
disks apparently only differ from their larger siblings by their small sizes.
The origin for such small sizes remains unknown,
and requires higher resolution observations with ALMA and the JVLA
to be unveiled. Such observations would also reveal the ability of these disks to form
planets.

\begin{acknowledgements}
We thank A.Bacmann for help with the V836 Tau data.
This work was supported by ``Programme National de Physique Stellaire'' (PNPS) and ``Programme
National de Physique Chimie du Milieu Interstellaire'' (PCMI) from INSU/CNRS.
This research has made use of the SIMBAD database,
operated at CDS, Strasbourg, France
\end{acknowledgements}

\bibliography{faint-disk-clean}
\bibliographystyle{aa}

\newpage
\appendix

\onecolumn

\section{Impact of Seeing}
\label{app:self}

\begin{figure*}[b!]
   \includegraphics[width=15.0cm]{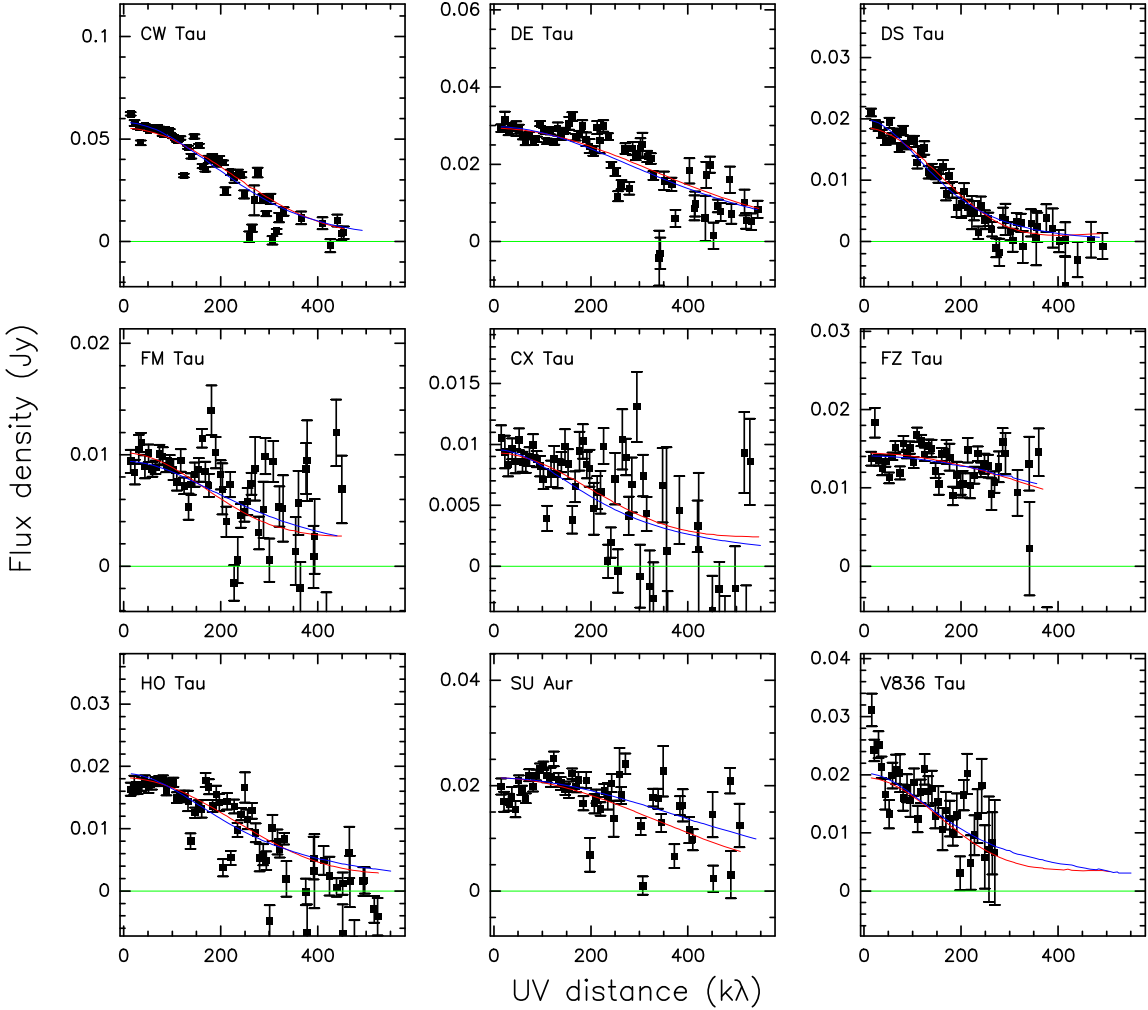}
 \caption{Deprojected visibility profiles for the sources on a common relative scale, with best-fit
 profiles superimposed. Partially self-calibrated version. Red is for the power-law fit, blue for the
 viscous profile.}
  \label{fig:cont}
\end{figure*}

Self-calibration is essential to measure the intrinsic source
  size, as the original phase noise on the longest baselines is
  relatively high. To illustrate this issue, we also used a data set
  in which only the most compact configurations (for which the
  correlated flux is well above the noise) have been
  self-calibrated. The most affected disk parameters are given in Table
  \ref{tab:cont} for fully and partially self-calibrated data. The
  corresponding (de-rotated and de-projected) visibilities are shown in
  Fig.\ref{fig:cont} (to be compared with Fig.\ref{fig:self}). The
  coherence loss for uv distances above 300 m is quite significant,
  allowing to derive only a lower limit to the disk sizes for the more
  compact ones. Self-calibration has no strong impact on the derived
  disk orientations, and only small one (due to seeing effects) on the
  inclination. As shown in Table \ref{tab:cont}, the recovered flux is
  slightly larger and the sources are more compact, requiring higher
  surface densities to reproduce the observations.

\begin{table*}[b!]
\caption{Effect of phase self-calibration}
\label{tab:cont}
\begin{tabular}{c |  cccc | cccc}
 Source        & \multicolumn{4}{c|}{Partially self-calibrated} & \multicolumn{4}{c}{Self-calibrated} \\
\hline
 &  Flux  &  Rout &  R$_c$ & $\Sigma_{10}$ &  Flux  &  Rout &  R$_c$ & $\Sigma_{10}$ \\
 & (mJy) &  (AU) &  (AU) & (g.cm$^{-2}$) & (mJy) &  (AU) &  (AU) & (g.cm$^{-2}$)  \\
\hline
\hline
FM Tau  &    9.7 $\pm$    0.2
 &   68 $\pm$  5
 &   33 $\pm$  4
 &       3
 &   11.3 $\pm$    0.2
 &   20 $\pm$  2
 &    3 $\pm$  4
 & 20
 \\
CW Tau &   55.8 $\pm$    0.3
 &   58 $\pm$  4
 &   30 $\pm$  2
 &      26
 &   58.7 $\pm$    0.4
 &   39 $\pm$  2
 &   18 $\pm$  2
 & 135
 \\
CX Tau  &    9.4 $\pm$    0.2
 &   30 $\pm$  8
 &   22 $\pm$  5
 &       4
 &    9.6 $\pm$    0.2
 &   17 $\pm$ 12
 &    8 $\pm$  2
 & 7
 \\
DE Tau  &   27.6 $\pm$    0.2
 &   36 $\pm$  1
 &   21 $\pm$  1
 &      16
 &   29.5 $\pm$    0.2
 &   20 $\pm$  8
 &    7 $\pm$  3
 & 300
 \\
FZ Tau  &   13.3 $\pm$    0.2
 &   28 $\pm$  3
 &   14 $\pm$  5
 &       9
 &   14.7 $\pm$    0.1
 &    9 $\pm$  3
 &    2 $\pm$  4
 & 20
 \\
HO Tau  &   17.0 $\pm$    0.2
 &   51 $\pm$  2
 &   37 $\pm$  1
 &       5
 &   17.7 $\pm$    0.2
 &   48 $\pm$  3
 &   22 $\pm$  1
 & 13
 \\
DS Tau  &   18.8 $\pm$    0.3
 &   72 $\pm$  4
 &   53 $\pm$  3
 &       6
 &   20.1 $\pm$    0.3
 &   70 $\pm$  4
 &   41 $\pm$  3
 & 13
 \\
SU Aur  &   18.2 $\pm$    0.3
 &   35 $\pm$  3
 &   21 $\pm$  2
 &       10
 &   22.5 $\pm$    0.2   
 &   11 $\pm$  5
 &    4 $\pm$  2
 &  3
 \\
V836 Tau  &   21.9 $\pm$    0.7
 &   75 $\pm$ 17
 &   49 $\pm$ 25
 &       4
 &   28.0 $\pm$    0.7
 &   48 $\pm$  9
 &   27 $\pm$  7
 & 15
 \\
\hline
\end{tabular}
\end{table*}

\section{Error bars derivation}
\label{app:errors}

The use of the covariance matrix to estimate the error bars is well suited
only for the most resolved sources. For the more compact ones, error bars on several
parameters are necessarily asymmetric. The upper bound on $R_\mathrm{out}$ is
determined by the visibility curves, while the lower bound is set by the necessity to
provide a sufficient total flux. $R_\mathrm{out}$ is thus well constrained even for
these very compact sources. The inclination of the very compact sources is of course
loosely defined, as it can only be constrained by measuring different sizes for different
orientations of the baselines. However, this has only a moderate impact on $R_\mathrm{out}$:
as the contribution of an optically thick core is in general
dominant, $R_\mathrm{out}^2 \cos{(i)}$ is to first order constant, unless
the disk becomes nearly edge-on.

\begin{figure}
   \includegraphics[width=6.0cm]{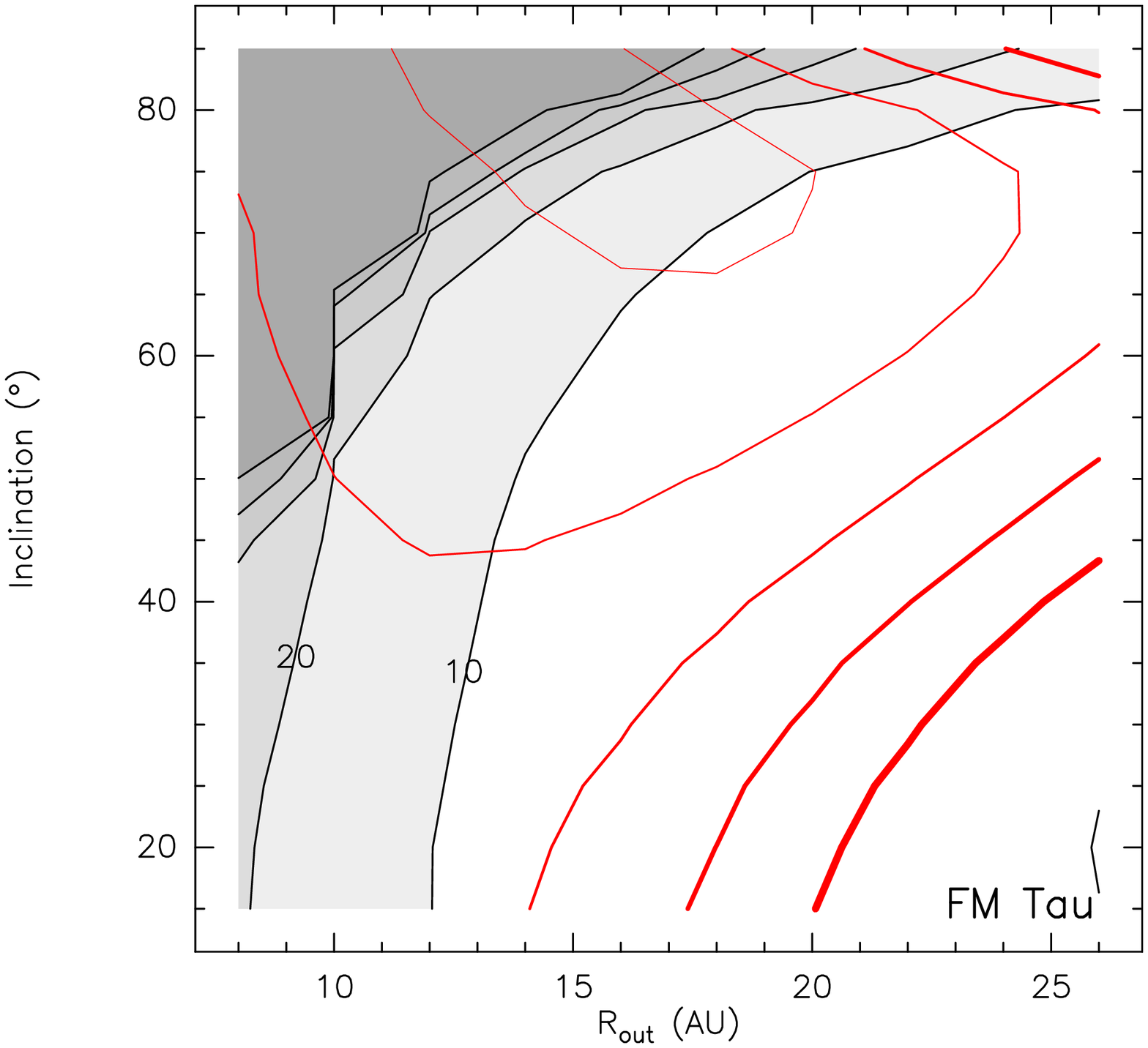}
   \includegraphics[width=6.0cm]{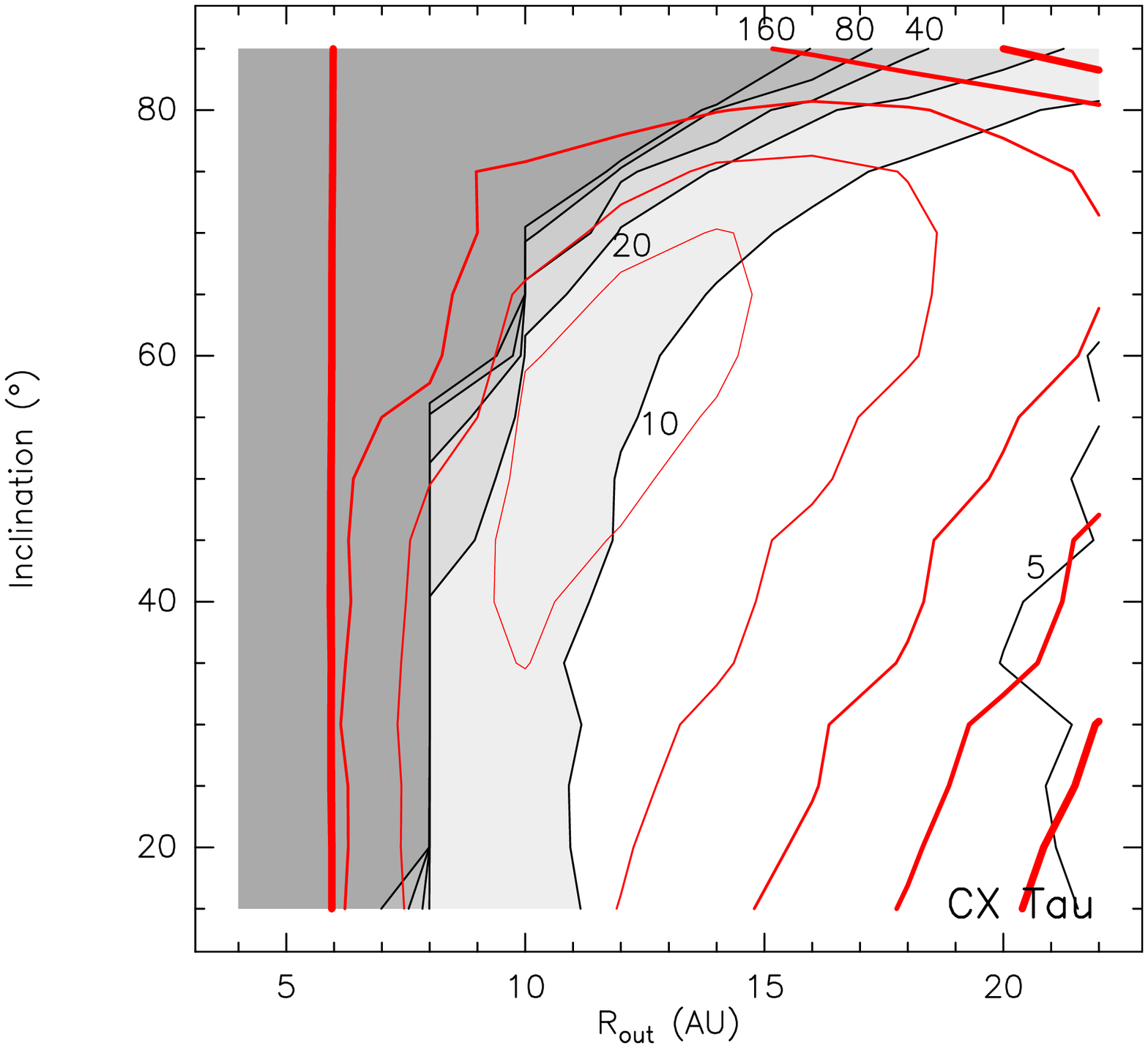}
   \includegraphics[width=6.0cm]{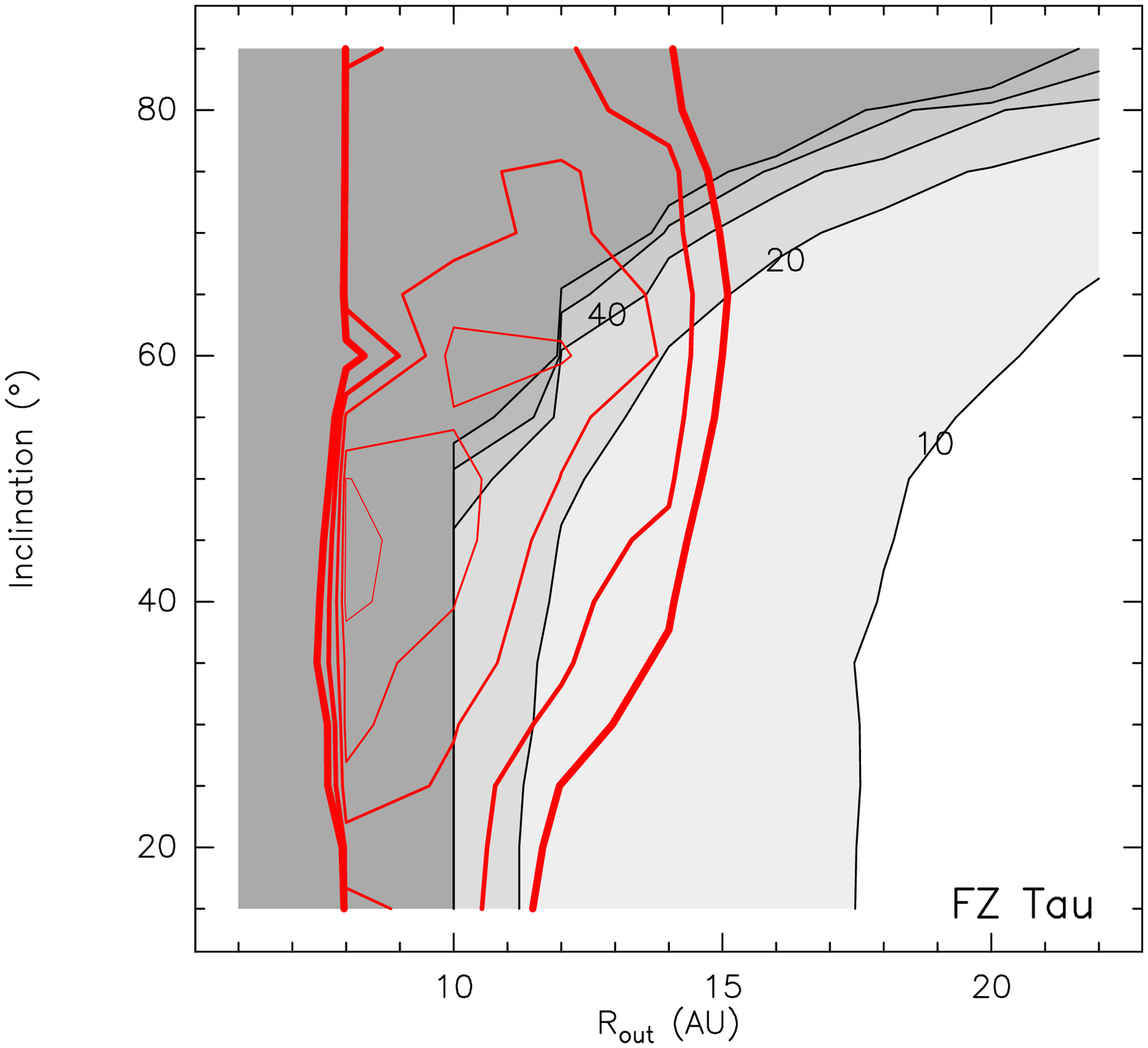}
   \includegraphics[width=6.0cm]{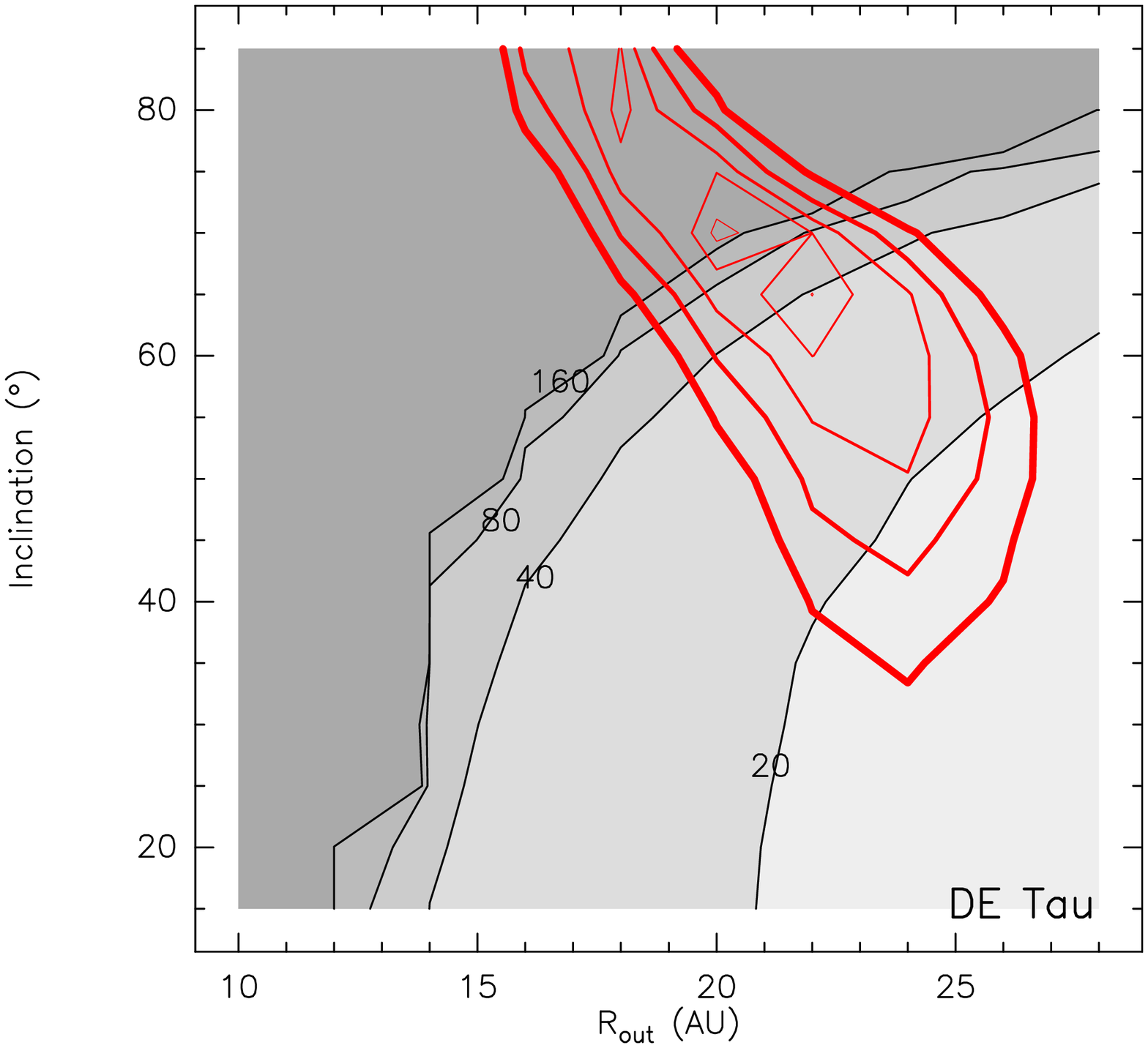}
   \includegraphics[width=6.0cm]{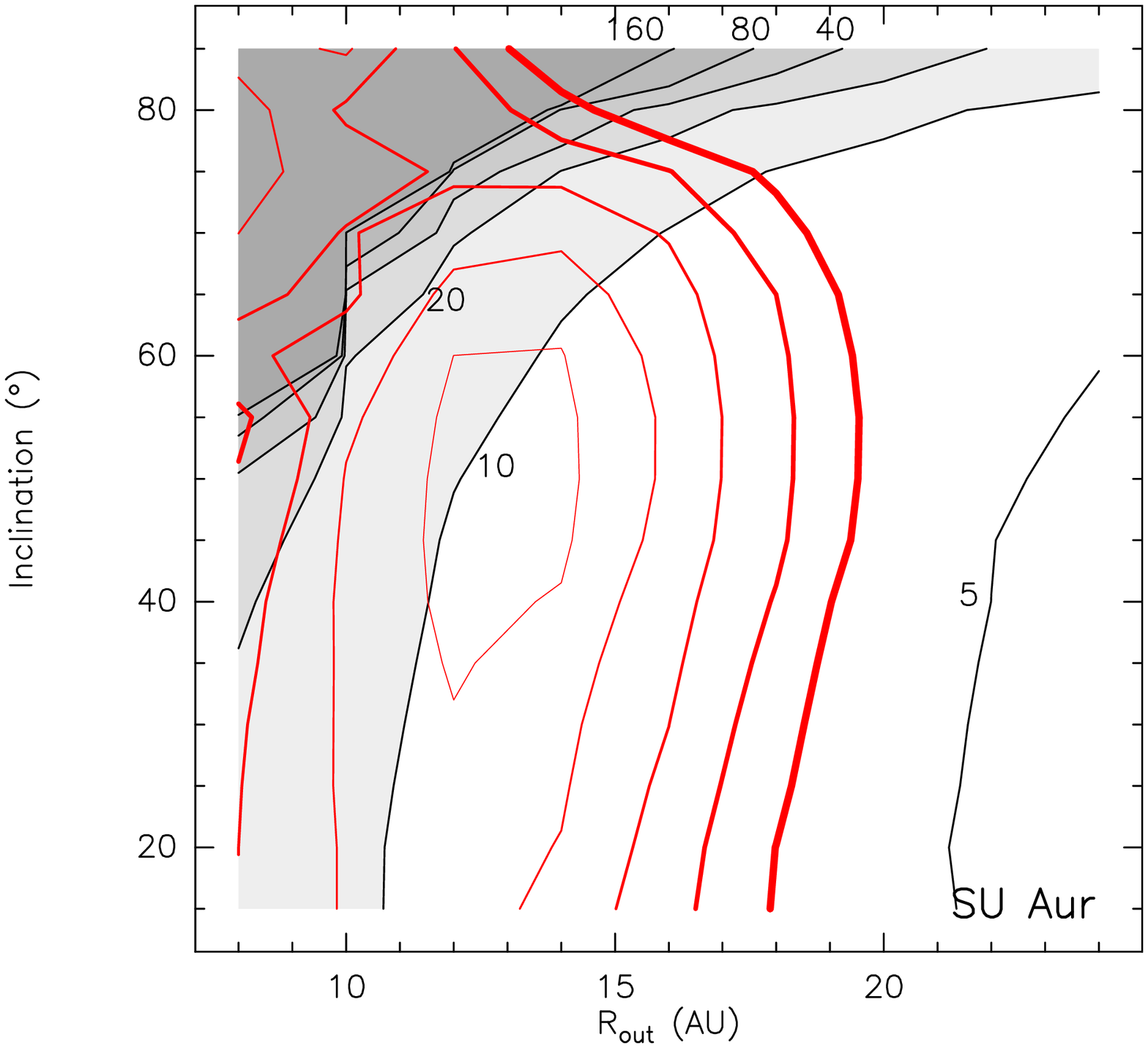}
 \caption{$\chi^2$ contours as a function of  $R_\mathrm{out}$ and $i$ for all
 compact sources (red contours,
 1 to 5 $\sigma$ by increasing width), overlaid on the surface density values in grey scale. The
 surface density contours are 5, 10, 20, 40, 80 and 160 g\,cm$^{-2}$ (of dust+gas, assuming a gas to dust
 ratio of 100). The source name is indicated in each panel.
 }
  \label{fig:chi2-whole}
\end{figure}

To avoid any substantial bias on the derivation of $R_\mathrm{out}$ and its
errorbars for small sources, we explored a 2-parameters
$\chi^2$ surface as a function of ($R_\mathrm{out}$,i), fitting position, orientation,
surface density and emissivity index.  The results are presented in Fig.\ref{fig:chi2-whole}.
These figures show the expected behaviour for $R_\mathrm{out}$.  The errors quoted in Table 
\ref{tab:self}
are in general conservative: the lower
bound on $R_\mathrm{out}$ is sharply constrained due to the flux requirement (see Eq.\ref{eq:fluxthick}).

Note that in practically all cases, the optically thick solution ($\Sigma > 50$ g\,cm$^{-2}$)
is within 2 $\sigma$ of the best fit value.  We also point out that given the unavoidable phase errors
even after self-calibration, the residual seeing effects increase the best fit radius. The quoted
best fit surface densities are thus lower limit in this respect, since only noise is included in the
errorbars.

\end{document}